\documentclass[12pt]{article}
\pdfoutput=1
\usepackage{amssymb,amsmath}
\usepackage{epsfig}
\usepackage{epstopdf}
\usepackage{latexsym}
\usepackage{graphicx}
\usepackage{subfigure}
\usepackage{booktabs}
\usepackage{bbm}
\usepackage[margin=20pt,small]{caption}

\usepackage[toc]{appendix}

\usepackage{color}
\usepackage{datetime}

\ifpdf
\fi

\renewcommand{\theequation}{\arabic{section}.\arabic{equation}}

\def\cN{{\cal N}}

\def\cL{{\cal L}}

\def\cS{{\cal S}}

\def\cL{{\cal L}}
\def\cA{{\cal A}}
\def\cB{{\cal B}}
\def\cV{{\cal V}}

\definecolor{cardinal}{rgb}{0.6,0,0}
\definecolor{darkgreen}{rgb}{0,0.5,0}
\definecolor{golden}{rgb}{0.92, 0.7, 0}
\definecolor{midnight}{rgb}{0, 0, 0.5}
\definecolor{darkblue}{rgb}{0.2, 0, 0.8}

\begin{document}  

\begin{titlepage}

\begin{flushright}
UTTG-02-12 \\
NSF-KITP-12-025
\end{flushright}
 
\bigskip
\bigskip
\bigskip
\bigskip
\centerline{\Large \bf Chiral Symmetry Breaking and External Fields}
\bigskip
\bigskip
\centerline{{\Large \bf in the}}
\bigskip
\bigskip
\centerline{{\Large \bf Kuperstein--Sonnenschein Model}}
\bigskip
\bigskip
\centerline{{\bf M. Sohaib Alam$^{1}$,  Vadim S. Kaplunovsky$^{1}$ and Arnab Kundu$^{1, 2}$}}
\bigskip
\centerline{$^1$Theory Group, Department of Physics}
\centerline{University of Texas at Austin} \centerline{Austin, TX 78712, USA.}
\bigskip
\centerline{$^2$ Kavli Institute for Theoretical Physics}
\centerline{University of California}
\centerline{Santa Barbara, CA 93106-4030, USA.}
\bigskip
\centerline{{\rm  malam@physics.utexas.edu, vadim@physics.utexas.edu, arnab@physics.utexas.edu} }
\bigskip
\bigskip

\begin{abstract}
\noindent A novel holographic model of chiral symmetry breaking has been proposed by Kuperstein and Sonnenschein by embedding non-supersymmetric probe D7 and anti-D7 branes in the Klebanov-Witten background. We study the dynamics of the probe flavours in this model in the presence of finite temperature and a constant electromagnetic field. In keeping with the weakly coupled field theory intuition, we find the magnetic field promotes spontaneous breaking of chiral symmetry whereas the electric field restores it. The former effect is universally known as the ``magnetic catalysis" in chiral symmetry breaking. In the presence of an electric field such a condensation is inhibited and a current flows. Thus we are faced with a steady-state situation rather than a system in equilibrium. We conjecture a definition of thermodynamic free energy for this steady-state phase and using this proposal we study the detailed phase structure when both electric and magnetic fields are present in two representative configurations: mutually perpendicular and parallel. 
\end{abstract}

\end{titlepage}

\newpage


\tableofcontents

\section{Introduction}

The gauge-gravity duality\cite{Maldacena:1997re, Gubser:1998bc, Witten:1998qj} (for a review, see \cite{Aharony:1999ti}) provides us with a remarkable tool to study a large class of strongly coupled large $N_c$ gauge theories. Within these class of theories, much effort has been spent trying to construct holographic models which share some of the key features of QCD at strong coupling, such as the confinement/deconfinement transition, chiral symmetry breaking and numerous other properties which are of recent phenomenological interests. See {\it e.g.} the review \cite{Gubser:2009md} for more details. The hope is any lesson learnt using these models will teach us useful lessons about QCD in some universal (and at least qualitative) sense.

In this article we will focus entirely on the physics of chiral symmetry breaking. In the holographic construction the fundamental matter fields are introduced by considering $N_f$ ``flavour branes" in the background of $N_c$ ``colour branes" and the global symmetry associated with these flavour branes is identified with the chiral symmetry. In an analogue of the quenched approximation, the problem simplifies in the probe limit where $N_f \ll N_c$ and thus the gravitational backreaction of the flavour branes can be safely ignored. The dynamics of the probe branes is then simply determined by the Dirac-Born-Infeld (DBI) action (supplemented by the Chern-Simons action when necessary) in the given gravitational background. This was initially done in \cite{Karch:2002sh} by considering probe ${\rm D7}$-branes in the background of $N_c$ ${\rm D3}$-branes. The background geometry there is given by ${\rm AdS}_5 \times S^5$ and the dual field theory is the $\cN=4$ super Yang-Mills. However, the global flavour symmetry in \cite{Karch:2002sh} is only a $U(1)$ and does not resemble the chiral symmetry group in QCD. Besides, the ${\rm D7}$-brane embeddings are $\frac{1}{2}$-BPS which necessarily implies that the chiral condensate identically vanishes and there is no spontaneous chiral symmetry breaking. We will refer to this as the ${\rm D3}-{\rm D7}$ model.\footnote{If we embed a probe D7-brane in a deformed confining D3-brane geometry ({\it e.g.} the Constable-Myers background), then the axial $U(1)$ (corresponding to the rotation in the directions transverse to the D7-brane) can be broken spontaneously\cite{Babington:2003vm}. This further allows one to compute the mass of the $\rho$-meson in terms of the mass of the $\pi$-meson, see {\it e.g.} \cite{Erdmenger:2007cm}.}

The Sakai-Sugimoto model\cite{Sakai:2004cn, Sakai:2005yt}, on the other hand, is based on considering ${\rm D8}$ and $\overline{\rm D8}$-branes in the non-extremal ${\rm D4}$-brane background. This brane--anti-brane pair is separated in the UV which gives rise to an $U(N_f)_L \times U(N_f)_R$ flavour symmetry, very similar to the chiral symmetry group in QCD. In the IR, these branes merge together smoothly spontaneously breaking the chiral symmetry to a diagonal $U(N_f)$. This model gives a simple and elegant geometric realization of the chiral symmetry breaking in QCD.

Recently a similar geometric mechanism of spontaneous chiral symmetry breaking has been introduced in \cite{Kuperstein:2008cq} by considering ${\rm D7}$/$\overline{{\rm D7}}$-branes in the conformal Klebanov-Witten background\cite{Klebanov:1998hh}. We will call this the Kuperstein-Sonnenschein model. The ${\rm D7}$/$\overline{{\rm D7}}$-branes wrap a three cycle in the internal manifold $T^{1,1} \cong S^2 \times S^3$ and is extended along the rest of the conifold $\mathbb{R}^+ \times S^2$. At zero temperature the brane--anti-brane pair has no choice but to dynamically join in the IR, which realizes spontaneous breaking of chiral symmetry: $U(N_f)_L \times U(N_f)_R \to U(N_f)_{\rm diag}$. Since the background is conformal, the two branches corresponding to the brane and the anti-brane produce an asymptotic angle separation of $\Delta\phi_\infty = (\sqrt{6}/4) \pi$ which is independent of the IR point where the brane--anti-brane pair joins. This asymptotic angle separation corresponds to the coupling of the corresponding operator introduced in the dual gauge theory.

It is worth remarking on the differences between this model and the Sakai-Sugimoto model. First, the Sakai-Sugimotmo model contains a running dilaton which diverges in the UV and one needs to worry about the UV completion of the theory. Second, the Sakai-Sugimoto model is built upon ${\rm D4}$-branes compactified on a spatial circle which is dual to a $(4+1)$-dimensional gauge theory at energies bigger than the compactification scale. The Kuperstein-Sonnenschein model avoids these two drawbacks rather simply: the dilaton does not run and by construction this is dual to an honest $(3+1)$-dimensional gauge theory. Thus the Kuperstein-Sonnenschein model has certain advantages over the Sakai-Sugimoto model.

Within the probe approximation, it is possible to further study the physics of chiral symmetry breaking in the presence of external parameters, such as temperature, constant electromagnetic field {\it etc}. In this article, we study the effect of finite temperature and a constant electromagnetic field in the Kuperstein-Sonnenschein model. Having a finite temperature corresponds to introducing a black hole in the bulk geometry. This corresponds to having the AdS-Schwarzschild$\times T^{1,1}$-background. The physics at finite temperature is rather simple because of the underlying conformal invariance. Since there is no other scale, chiral symmetry is restored as soon as any temperature is turned on.

At finite temperature, we introduce a constant electromagnetic field by exciting gauge fields on the worldvolume of the probe brane. This gauge field, in the probe limit, does not modify the background; it only affects the probe. Thus the electromagnetic field we consider couples only to the flavour sector and results in a non-trivial phase structure for the flavours. Recall that at zero temperature and vanishing external fields the coupling in the dual field theory $\Delta \phi_\infty$ has a fixed value. When we introduce these external parameters we do not insist that the coupling remain fixed at this value. If we have well-defined UV theory for a given coupling, then changing this coupling would imply we change the theory as well, which may not be desirable.\footnote{We thank Anatoly Dymarsky for raising this point.} However, as pointed out in \cite{Kuperstein:2008cq} the operator corresponding to $\Delta \phi_\infty$ in the dual field theory is not completely understood. Our approach, thus, is entirely guided by the holographic construction and is more in the spirit of condensed matter physics where one allows various couplings in the theory to depend on the external parameters introduced in the system and scans the space of possible phases as the couplings change.

Phase diagrams with similar external parameters have been studied in the ${\rm D3}-{\rm D7}$ model in \cite{Filev:2007gb, Albash:2007bk, Albash:2007bq, Erdmenger:2007bn, Evans:2010xs} and in the Sakai-Sugimoto model in \cite{Bergman:2008sg, Johnson:2008vna}; for a comparative account of these studies see {\it e.g.} \cite{Kundu:2010ye}.\footnote{For more recent studies involving the ${\rm D3}-{\rm D7}$ model, see {\it e.g.} \cite{Evans:2011mu, Evans:2011tk}.} It has been found that the magnetic field promotes the spontaneous breaking of the global flavour symmetry and results in a non-trivial phase diagram in the temperature vs magnetic field plane. This effect is widely recognized as the {\it magnetic catalysis} in chiral symmetry breaking\cite{Gusynin:1995nb, Semenoff:1999xv, Miransky:2002eb}. The key physics behind this phenomenon is an effective dimensional reduction of the problem in the presence of a magnetic field. In a strong magnetic field, the lowest Landau level plays an important role and reduces the dynamics from $d$-spatial dimensions to $(d-2)$-spatial dimensions.  From the holographic point of view, this catalysis effect is seen as a magnetic field-induced bending of the probe flavour brane. An electric field on the other hand favours symmetry restoration and drives a current\cite{Karch:2007pd, Albash:2007bq, Bergman:2008sg}. This current is non-zero even in the absence of finite chemical potential or charge density. The key physics behind this is simple: charge carriers are created from the vacuum via thermal and quantum fluctuations. The holographic realization of this effect is rather elegant. In the presence of a constant electric field, the probe brane excites an appropriate gauge field on its worldvolume which is dual to a boundary current carried by the fundamental flavours. In the T-dual picture, having a constant electric field on the worldvolume of the probe brane is equivalent to considering the brane with some angular velocity, as considered in {\it e.g.} \cite{Albash:2006bs, Das:2010yw}. In such a case, due to gravitational red-shift, the local speed of propagation on the probe brane can exceed the speed of light near the infrared region of the bulk geometry. To prevent such superluminal propagation the probe brane can develop a non-trivial profile along another transverse direction. When we T-dualize back to our original configuration, this ``extra" profile maps to a gauge field living on the worldvolume of the D7-brane which is holographically dual to a current in the boundary theory.

In this article we demonstrate that a similar {\it magnetic catalysis} effect exists in the Kuperstein-Sonnenschein model. At vanishing magnetic field, the finite temperature immediately restores the symmetry without undergoing any phase transition. At non-zero magnetic field, there is a first order phase transition at some critical temperature below which chiral symmetry is broken and beyond which it is restored. This happens at a critical value of the coupling $\Delta \phi_\infty$ in the boundary theory. As we increase the magnetic field, the critical coupling increases and at infinitely large magnetic field approaches a finite constant value. We also study the thermodynamics associated with this first order phase transition.

The thermal physics in the presence of an electric field is more subtle. The presence of a current in the boundary theory implies that we are dealing with a steady-state system rather than an equilibrium system. The identification of a thermodynamic free energy and hence to determine the corresponding phase diagram in this case becomes more subtle. Previous works in {\it e.g.} \cite{Albash:2007bq, Bergman:2008sg} have made use of a ``Maxwell construction" to determine the phase transition point. However, we believe this is inappropriate. The ``insulating" phase has vanishing current and the ``conducting" phase has a non-zero current. The current jumps to a constant non-vanishing value across the phase transition, but we do not see the ``metastable" states where the current smoothly interpolates between zero and the non-vanishing constant value across the phase transition. To count the energetics properly, the Maxwell construction relies on the presence of these metastable states. We circumvent this issue by proposing a definition of the thermodynamic free energy in the conducting phase in terms of the probe brane's on-shell action. The prescription is: we first need to supplement the usual DBI piece with a boundary term in order to have a well-defined variational problem. Then we need to put an IR cut-off at a radial position which we call the ``pseudo-horizon" that emerges as a natural radial scale in the problem. We argue that this cut-off is natural since the open string degrees of freedom effectively see a horizon at this position.

Using our proposal of the thermodynamic free energy we then explore the rich phase diagrams when both electric and magnetic fields are present. We choose two representative configurations: perpendicular electric and magnetic fields and parallel electric and magnetic fields. In both these cases the qualitative features of the phase diagrams are similar and conforms to our general intuition of temperature and electric field favouring chiral symmetry restoration and a magnetic field promoting symmetry breaking. We also argue that when both electric and magnetic fields are present, the corresponding phase diagrams have non-trivial structure only when the electric field is smaller than the magnetic field. In the regime where the electric field is greater than the magnetic field, we do not have any chiral symmetry broken phase.

This paper is organized as follows: We briefly review the Kuperstein-Sonnenschein model in section 2. In section 3, we briefly discuss the physics at finite temperature. We introduce a magnetic field in section 4 and in section 5 we study the effect of both temperature and magnetic field. In section 6, we introduce an electric field and discuss the subtleties associated in identifying a free energy and conjecture a proposal to do sensible thermodynamics. We use this proposal in section 7 to study the detailed phase structure in the presence of both perpendicular and parallel electric and magnetic fields. Finally we conclude in section 8 with open questions and future directions. Some relevant details have been relegated to three appendices.

\section{The Kuperstein-Sonnenschein Model} \label{T0}

Let us begin by briefly reviewing the Kuperstein-Sonnenschein model
introduced in \cite{Kuperstein:2008cq}.
We start with the $AdS_5 \times T^{1,1}$ background (first obtained in \cite{Romans:1984an} and then explored in the context of AdS/CFT in \cite{Klebanov:1998hh})
which is the near-horizon limit of a stack of $N_c$ D3-branes placed on the tip of a conifold.
The metric is
\begin{eqnarray}\label{kw}
ds^2 & = & \frac{r^2}{R^2} dx_\mu dx^{\mu} + \frac{R^2}{r^2} ds_6^2 \,,\\
ds_6^2 & = & dr^2 + r^2 ds_{T^{1,1}}^2 \\
& = & dr^2 + \frac{r^2}{3} \left(
	\frac{1}{4} \left(f_1^2 + f_2^2 \right) + \frac{1}{3} f_3^2
	+ \left(d\theta - \frac{1}{2} f_2\right)^2
	+ \left(\sin\theta d\phi - \frac{1}{2} f_1\right)^2
	\right) \,,\nonumber
\end{eqnarray}
the dilaton is constant, and there is a self-dual 5-form RR flux
\begin{equation}
F_5\ =\ {4r^3\over g_s R^4}\,dr\wedge d^4 x\ 
- \ {R^4\over 27 g_s}\,\sin\theta\,d\theta\wedge d\phi\wedge f_1\wedge f_2\wedge f_3\,.
\end{equation}
In our notations 
$x^\mu$ are the four Minkowski directions, $r$ is the AdS-radial coordinate,
and $(f_1,f_2,f_3, \theta, \phi)$ represent the $T^{1,1}$ as a local
$S^3 \times S^2$ trivialization ---
the $f_{1,2,3}$ are unit differentials on the $S^3$ while $\theta$ and $\phi$
are spherical coordinates on the $S^2$.
Furthermore, $R$ is the AdS radius --- which obtains as
\begin{equation}
R^4\ =\ \frac{27\pi}{4}\, N_c g_s \alpha^{\prime2}\ =\ \lambda \alpha^{\prime2}
\end{equation}
where $g_s$ is the string coupling, $2\pi\alpha'$ is the inverse string tension,
and $\lambda$ is the 't~Hooft coupling of the dual 4D gauge theory;
later in this paper we shall use a related coupling
\begin{equation} \label{nlambda}
\bar\lambda\ =\ \frac{\pi^2}{4}\, \lambda \ . 
\end{equation}

The field theory dual to this background was constructed in \cite{Klebanov:1998hh}: it is an ${\cal{N}} =1$ superconformal quiver gauge theory with a gauge group $SU(N_c) \times SU(N_c)$ and  two bi-fundamental chiral superfields usually denoted by $A_{1,2}$, $B_{1,2}$. These fields transform in the $(N_c, \bar{N_c})$ and $(\bar{N_c}, N_c)$ representations of the gauge group $SU(N_c) \times SU(N_c)$. This theory has a further global $SU(2) \times SU(2) \times U(1)_{\rm R}$ symmetry. Under these two $SU(2)$ symmetries the bi-fundamentals transform as a doublet of one of the $SU(2)$'s and as a singlet of the other one.

Following \cite{Kuperstein:2008cq} we place the D7 and anti D7-brane along the Minkowski and the $S^3$-directions, and restrict to the equatorial embedding denoted by $\theta = \pi/2$, $\phi = \phi(r)$. The brane--anti-brane pair is separated in the $\phi$-direction at the UV boundary (at $r\to \infty$). This configuration preserves one of the global $SU(2)$'s of the background. The corresponding DBI action is given by
\begin{eqnarray}\label{kwact}
&& S = - \tau_{7} \int d^8 \xi \sqrt{- {\rm det} P[G]} = - {\cal N} \int dt dr r^3 \left[1 + \frac{r^2}{6} \left(\phi'\right)^2 \right]^{1/2} = - {\cal N} \int dt dr {\cal L} \ , \\
&& {\cal N} = \tau_{7} V_{\mathbb R^3} \frac{8\pi^2}{9} \ .
\end{eqnarray}
In the above equation, $\tau_{7}$ denotes the tension of the D7-brane, $\xi$ denotes the D7-brane worldvolume coordinates, $P[G]$ denotes the pull back of the background metric on the probe brane, ${\cal L}$ is the Lagrangian density. Here $V_{\mathbb R^3}$ is the volume of the spatial $\mathbb{R}^3$. 

The equation of motion resulting from the action in (\ref{kwact}) is given by
\begin{eqnarray}\label{kwLag}
\frac{(r^5/6) \phi'}{\left(1 + \frac{r^2}{6} (\phi')^2\right)^{1/2}} = c  \ , 
\end{eqnarray}
where $c$ is the constant of motion. The large $r$ behaviour of the profile is
\begin{eqnarray} \label{phi@as}
\phi (r) = \frac{\Delta \phi_\infty}{2} - \frac{3 c}{2 r^4} + \ldots \ ,
\end{eqnarray}
where $\Delta \phi_\infty$ is the asymptotic angle separation between the brane--anti-brane pair. It is clear from the asymptotic behaviour of the profile function that $\Delta\phi_\infty$ is the non-normalizable mode (corresponding to source/coupling in the boundary theory) and $c$ is the normalizable mode (corresponding to VEV/condensate in the boundary theory).

We can integrate the equation of motion in (\ref{kwLag}) analytically and the full solution is given by\cite{Kuperstein:2008cq}
\begin{eqnarray}\label{sol0}
\cos \left(\frac{4}{\sqrt{6}} \phi(r) \right) = \left(\frac{r_0}{r}\right)^4 \ , \quad {\rm with} \quad \phi'(r_0) \to \infty \quad \implies c = \frac{r_0^4}{\sqrt{6}} \ .
\end{eqnarray}
The boundary condition $\phi'(r_0) \to \infty$ ensures that the brane--anti-brane smoothly join at $r_0$. We have two branches of solutions with $\phi \in [0, \pi/2]$ and $\phi \in[-\pi/2 , 0]$. The first branch corresponds to the D7-brane and the second one to the $\overline{{\rm D7}}$-brane. As $r \to \infty$, we see that $\Delta \phi_{\infty} = \frac{\sqrt{6}}{4} \pi$. Thus, in fact, one gets a family of solutions (with the same asymptotic angle separation) parametrized by $r_0$ where the brane--anti-brane pair joins. For future references we call these profiles as the ``U-shaped" embeddings. It is easy to see that since there is no natural place for the brane--anti-brane pair to end separately, they must join together. We will see at finite temperature this is not the case any more. There is a special solution for $r_0 =0$ given by $\phi_{\pm} = \pm (\sqrt{6}/8) \pi$. Each branch of the family of solutions with $r_0 \not = 0$ is non-holomorphic and thus breaks supersymmetry completely. The two branches of the solution with $r_0 =0$ also break supersymmetry completely since they are not antipodal\cite{Kuperstein:2008cq}. 
\begin{figure}[htp]
\begin{center}
 {\includegraphics[angle=0,
width=0.65\textwidth]{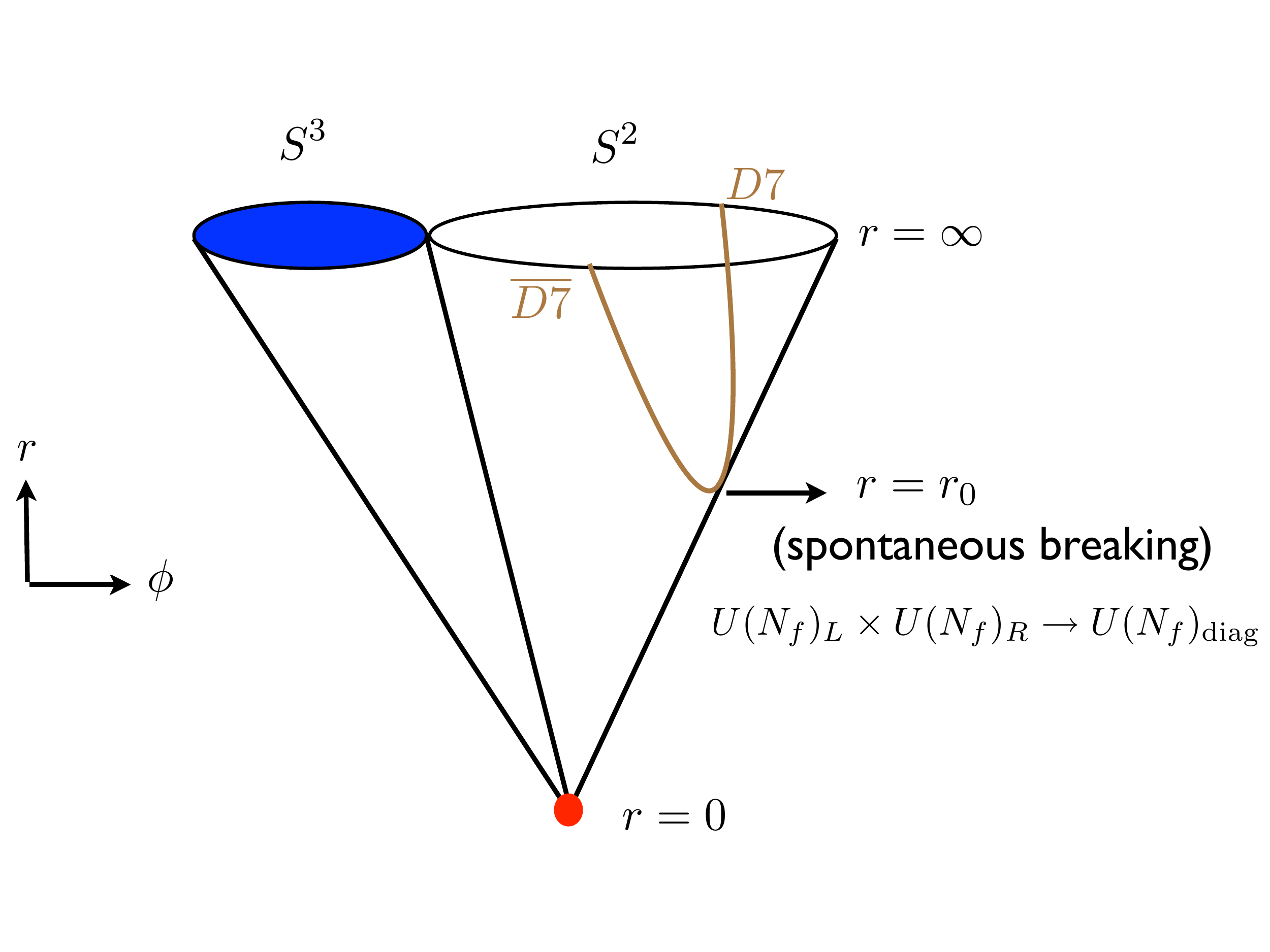}}
\caption{\small A schematic diagram showing the shape of the profile (in grey) of the brane--anti-brane pair. The red dot located at $r=0$ represents the conifold singularity. The brane--anti-brane pair joins at $r=r_0$ realizing the spontaneous breaking of chiral symmetry.}
\label{kwt0}
\end{center}
\end{figure}
The qualitative shapes of the brane--anti-brane profile have been demonstrated in fig.~\ref{kwt0}.

Introducing probe ${\rm D7}$ and $\overline{{\rm D7}}$ implies that we have introduced matter fields in the fundamental representation in the dual gauge theory. As argued in \cite{Kuperstein:2008cq}, adding the ${\rm D7}$/$\overline{{\rm D7}}$-brane corresponds to introducing left-handed/right-handed Weyl fermions in the dual gauge theory. Thus, in the UV where the ${\rm D7}$ and the $\overline{{\rm D7}}$ are separate, we have a global $U(N_f)_L \times U(N_f)_R$ flavour symmetry, where $N_f$ is the number of flavours.\footnote{Strictly speaking, here we take $N_f=1$.} This global flavour symmetry is dynamically broken to a diagonal $U(N_f)$ in the infrared where the brane--anti-brane pair joins. The asymptotic angle separation $\Delta \phi_{\infty}$ corresponds to the coupling of an operator in the dual gauge theory.\footnote{It is not completely clear at present what the corresponding operator is; in fact the quiver diagram of this theory (after introducing the fundamental matter) is not completely understood. Part of the complications arise from breaking supersymmetry completely, which means we can no longer use the technology of supersymmetric field theories to ``fix" various terms in the Lagrangian. Some thoughts and proposals on this are given in \cite{Kuperstein:2008cq}.} One would be tempted to identify the constant $c$ with the quark condensate corresponding to the breaking of the chiral symmetry, however since the corresponding operator is not well-understood at this moment we will make no such precise claim. Nonetheless, it is fair to say that the constant $c$ serves the purpose of an order parameter for the breaking of the chiral symmetry.

\section{Introducing Finite Temperature}

Let us now discuss the physics at finite temperature. The finite temperature background is given by $AdS_5$-Schwarzschild$\times T^{1,1}$. Also, we need to Euclideanize the time direction and periodically identify along a circle. The temperature is then simply given by the inverse period. In Euclidean signature this background is explicitly given by
\begin{eqnarray}\label{KW@T}
ds^2 = \frac{r^2}{R^2} \left( f(r) dt_E^2 + dx_i^2 \right) + \frac{R^2}{r^2} \frac{dr^2}{f(r)} + R^2 ds_{T^{1,1}}^2 \ , \quad f(r) = 1 - \left(\frac{r_H}{r}\right)^4 \ ,
\end{eqnarray}
where $t_E$ is the Euclidean time direction, $x_i$ with $i = 1, 2, 3$ represent the spatial 3-directions, $r_H$ is the location of the horizon and the temperature is given by $T = r_H / (\pi R^2)$. Furthermore, $ds_{T^{1,1}}^2$ represents the metric on the $T^{1,1}$ which is given in (\ref{kw}). This background corresponds to the phase of the dual gauge theory where the adjoint matter is deconfined.

\begin{figure}[htp]
\begin{center}
 {\includegraphics[angle=0,
width=0.65\textwidth]{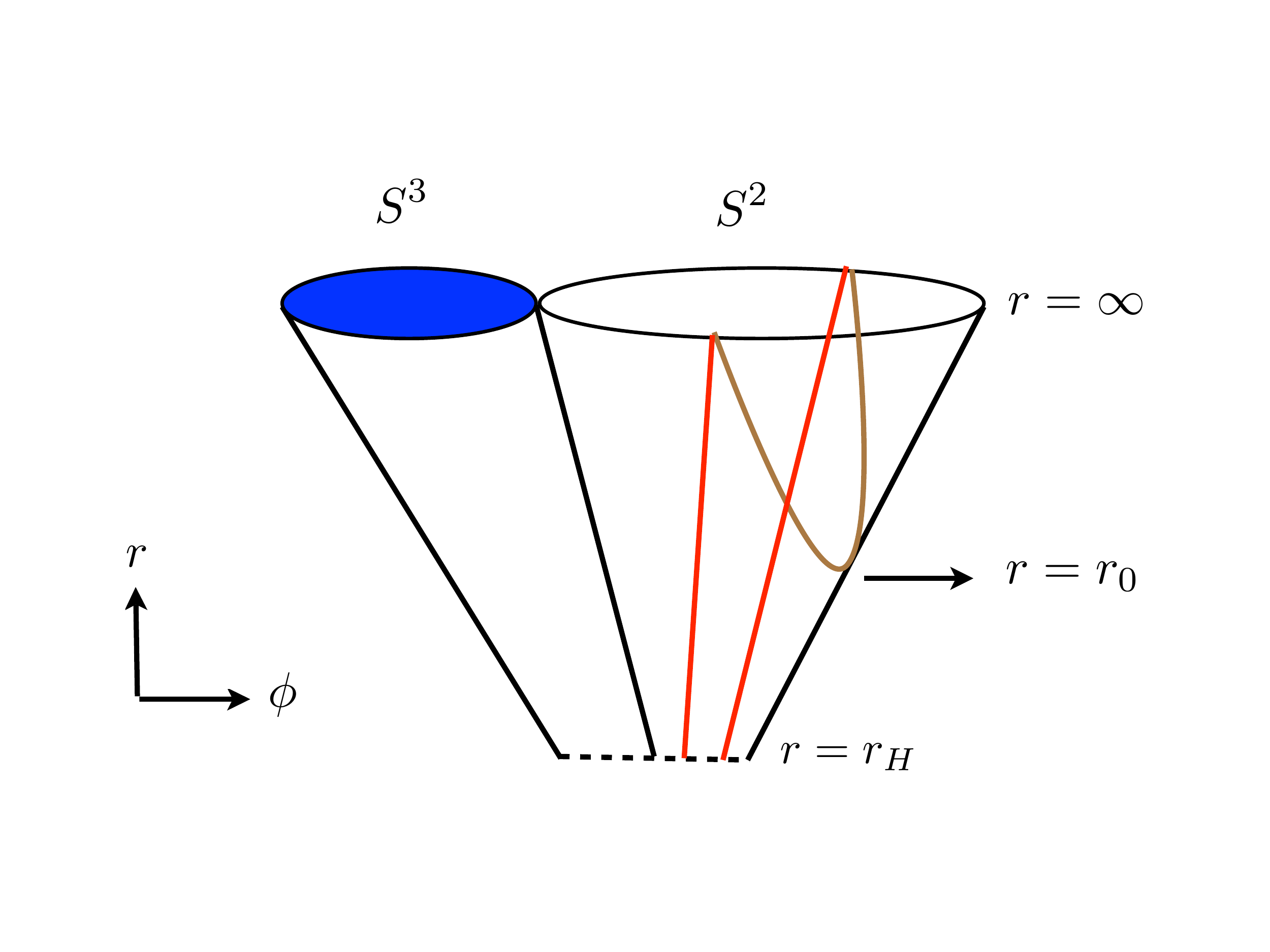}}
\caption{\small A schematic diagram showing the qualitative shapes of the probe brane--anti-brane pair when a non-zero temperature has been introduced. We have two classes of embeddings: the U-shaped ones (in grey) and the parallel ones (in red). The singularity of the cone is hidden behind the horizon located at $r=r_H$.}
\label{kwt}
\end{center}
\end{figure}

Now, we introduce the ${\rm D7}$ and $\overline{{\rm D7}}$ pair along the same directions as in the zero temperature case. In this case the DBI action\footnote{Note that there is a relative -ve sign between the DBI action at finite temperature and the one at zero temperature. This simply stems from the fact that in the finite temperature case we are working in an Euclidean signature.} is given by
\begin{eqnarray}\label{kwT}
S & = & \tau_{7} \int d^8 \xi \sqrt{ {\rm det} P[G]} = {\cal N}_T \int dr r^3 \left( 1 + \frac{r^2}{6} f (\phi')^2 \right)^{1/2} \ , \\
{\cal N}_T & = & \frac{{\cal N}}{T}  \ .
\end{eqnarray}
Note that the definition of ${\cal N}_T$ in this case differs from the zero temperature case by a factor of the temperature. The resulting equation of motion is given by
\begin{eqnarray}\label{eomT}
\frac{(r^5/6) f \phi'}{\left(1 + \frac{r^2}{6} f (\phi')^2\right)^{1/2}} = c  \ ,
\end{eqnarray}
where $c$ is the constant of motion. The equation (\ref{eomT}) is not analytically solvable. There are two possible classes of solutions to the equation (\ref{eomT}): the U-shaped ones for which we have $c = \frac{r_0^4}{\sqrt{6}} f(r_0)^{1/2}$, where $r_0$ is the point where the brane--anti-brane pair smoothly joins; the second class of solutions are the ones where the brane and the anti-brane separately end on the horizon. These are given by $\phi_{\pm} = \pm {\rm const}$ (corresponding to $c=0$) which we henceforth call the ``$\parallel$ embedding". The $\pm$ sign corresponds to the ${\rm D7}$ and the $\overline{{\rm D7}}$-brane respectively. As before, the U-shaped embeddings correspond to spontaneous breaking of chiral symmetry and the $\parallel$ embeddings correspond to chiral symmetry restoration. Clearly $r_0$ and $r$ both have the following range: $r_H \le r_0, r \le \infty$. In fig.~\ref{kwt} we have pictorially demonstrated various possible profiles.

In a situation like this, one would typically expect as we vary the temperature the system undergoes a first order phase transition and at some critical temperature chiral symmetry is restored. However, we started with a conformal background and so far temperature is the only scale in the system. We do not have any other scale in terms of which such a critical temperature can be measured.\footnote{Note that $r_0$ where the brane--anti-brane pair joins seems to provide another scale in the system. However, in reality this is a modulus of the problem and this modulus is only perceived as the dimensionless asymptotic angle separation at the boundary, but not a scale in the system.} The conclusion therefore must simply be: If there exists a temperature that can restore the chiral symmetry, then the $\parallel$ embeddings will always be energetically favoured.

\begin{figure}[htp]
\begin{center}
 {\includegraphics[angle=0,
width=0.65\textwidth]{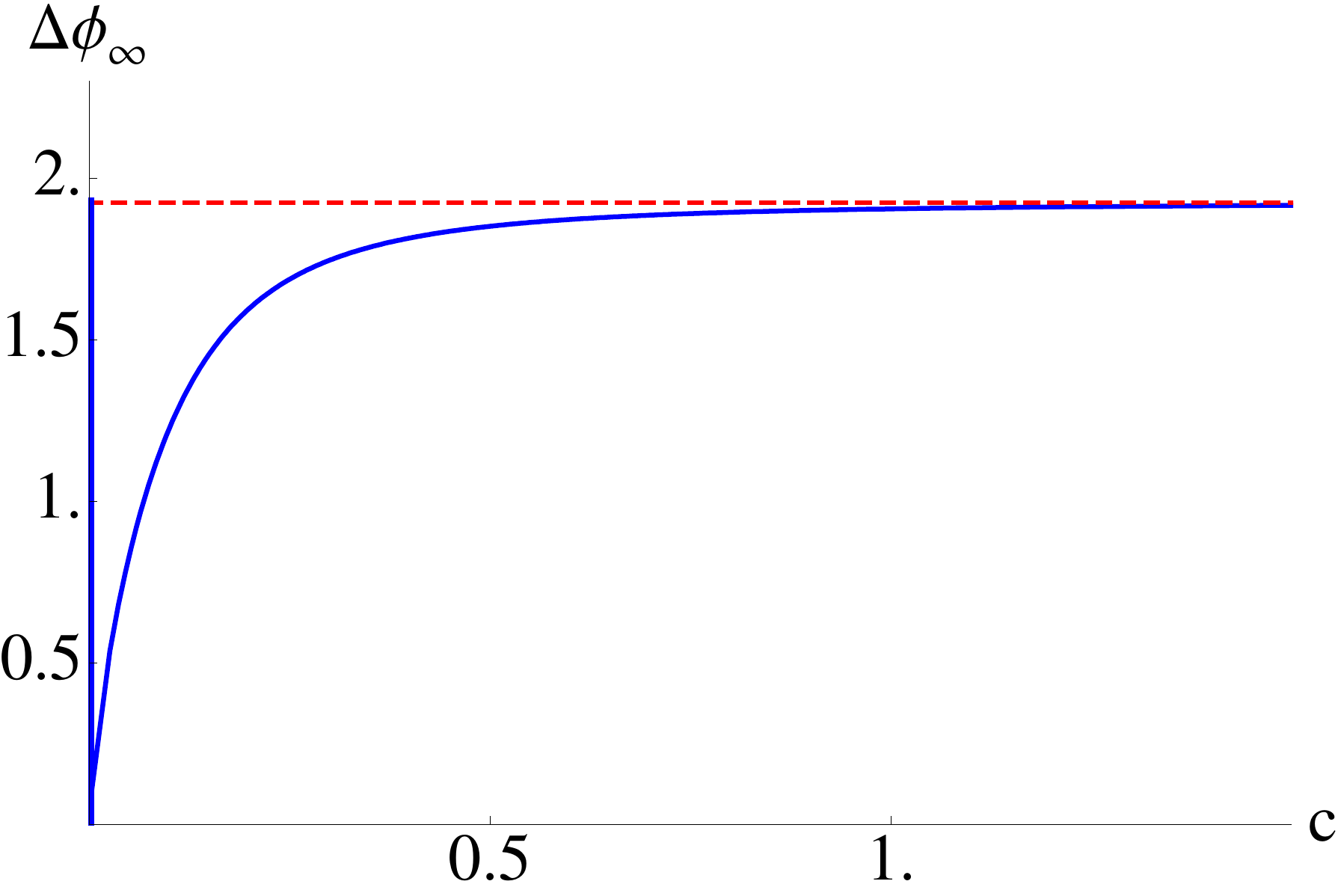}}
\caption{\small $\Delta \phi_\infty$ as a function of $c$. The red dashed line is the asymptotic boundary value at zero temperature. The blue curves (including the vertical line at $c = 0$) correspond to the space of solutions.}
\label{bc@c}
\end{center}
\end{figure}

Let us elaborate a bit more on this issue. As we have seen, the chiral symmetry broken phase measures a non-zero value of $c$ and the chiral symmetry restored phase has $c=0$. Thus the constant $c$ serves the purpose of an order parameter of this symmetry breaking, although it should not be confused with the chiral condensate. Thus we have the canonically conjugate variables $\{\Delta\phi_\infty, c \}$ and we should be able to see the signature of a first order phase transition (if it exists) in this plane. A numerical plot is shown in Fig.~\ref{bc@c}. 
To have a first order phase transition we would expect to see a turn-around behaviour of $\Delta\phi_\infty$ as $c$ is increased. However, here $\Delta\phi_\infty$ only approaches the value $(\sqrt{6}/4)\pi$ from below as $c \to \infty$. Hence we can conclude that this system does not have any first order phase transition.

To conclusively decide which embedding is favoured, we consider the following energy difference
\begin{eqnarray}
\Delta S = S_{U} - S_{\parallel} =  {\cal N}_T r_0^4 \left( \int_1^{\infty} dy y^3 \left[\left( 1+ \frac{f(1)}{y^8 f(y) - f(1)}\right)^{1/2} - 1 \right] -  \int_{y_H}^1 dy y^3 \right) \ ,
\end{eqnarray}
where we have defined
\begin{eqnarray}
y = \frac{r}{r_0} \ , \quad y_H = \frac{r_H}{r_0} \ , \quad f(y) = 1 - \left(\frac{y_H}{y}\right)^4 \ .
\end{eqnarray}
Now it is sufficient to check the sign of $\Delta S$ for any value of $y_H$, since every temperature is identical. We can perform a Taylor expansion of the integrand in the limit $y_H \to 0_+$ and at the leading order in $y_H$ obtain 
\begin{eqnarray}
\Delta S = ({\cal N_T} r_0^4 ) \frac{y_H^4} {8}  > 0 \ .
\end{eqnarray}
This clearly implies that the $\parallel$ embeddings are favoured. Thus finite temperature restores chiral symmetry.

\section{Introducing a Magnetic Field} \label{TB}

Let us first discuss the case of vanishing temperature. The relevant background is given in (\ref{kw}). Now we want to introduce a constant magnetic field on the worldvolume of the probe ${\rm D7}$ and $\overline{{\rm D7}}$-brane. Recall that the DBI action is given by
\begin{eqnarray}
S = - \tau_{7} \int d^8 \xi \sqrt{- {\det} \left (P [G + B] + (2 \pi \alpha' F \right)} \ ,
\end{eqnarray}
where $B$ is the background NS-NS field (which is zero in this case) and $F$ is the electromagnetic 2-form on the worldvolume of the probe brane.

Here we want to introduce a Minkowski gauge field, specifically a constant magnetic field, on the probe brane worldvolume. This can be achieved by simply exciting a gauge field of the form:\footnote{It can be checked {\it a posteriori} that this ansatz for the gauge field does satisfy the equations of motion resulting from the DBI action itself.} $A_ 3 = H x^2$ which gives a constant field strength $F_{23} = H$. This corresponds to having a constant magnetic field along the $x^1$-direction on the probe brane worldvolume. Since we are in the probe limit, this gauge field does not affect the 10-dimensional background. Thus in the dual field theory the adjoint matter is insensitive to this external field and only the fundamental matter couples to it. Our purpose here will be to investigate the effect of this constant field on the physics of chiral symmetry breaking.

With this gauge field, the action for the ${\rm D7}$/$\overline{{\rm D7}}$ is given by\footnote{It can be checked that there is no contribution coming from the Chern-Simons term.}
\begin{eqnarray}
S & = & - {\cal{N}} \int dt dr {\cal L} = - {\cal{N}} \int dt dr r^3 \left( 1 + \frac{r^2}{6} (\phi')^2 \right)^{1/2} \left( 1+ \frac{h^2}{r^4} \right)^{1/2} \ , \\
h & = & 2 \pi \alpha' R^2 H \ , \quad {\cal N} = \tau_{7} V_{\mathbb R^3} \frac{8\pi^2}{9} \ .
\end{eqnarray}
Introducing the magnetic field introduces a scale in the theory which is denoted by $h$. Thus we break conformal invariance explicitly even in the zero temperature case. The equation of motion resulting from this action is given by
\begin{eqnarray} \label{eomh}
\frac{r^3 \left( 1 + h^2/ r^4 \right)^{1/2} (r^2 /6) \phi'}{\left( 1 + (r^2/6) (\phi')^2 \right)^{1/2}} = c \ ,
\end{eqnarray}
where $c$ is the constant of motion. The asymptotic behaviour of the profile $\phi(r)$ is the same as given in (\ref{phi@as}).

\begin{figure}[!ht]
\begin{center}
 {\includegraphics[angle=0,
width=0.65\textwidth]{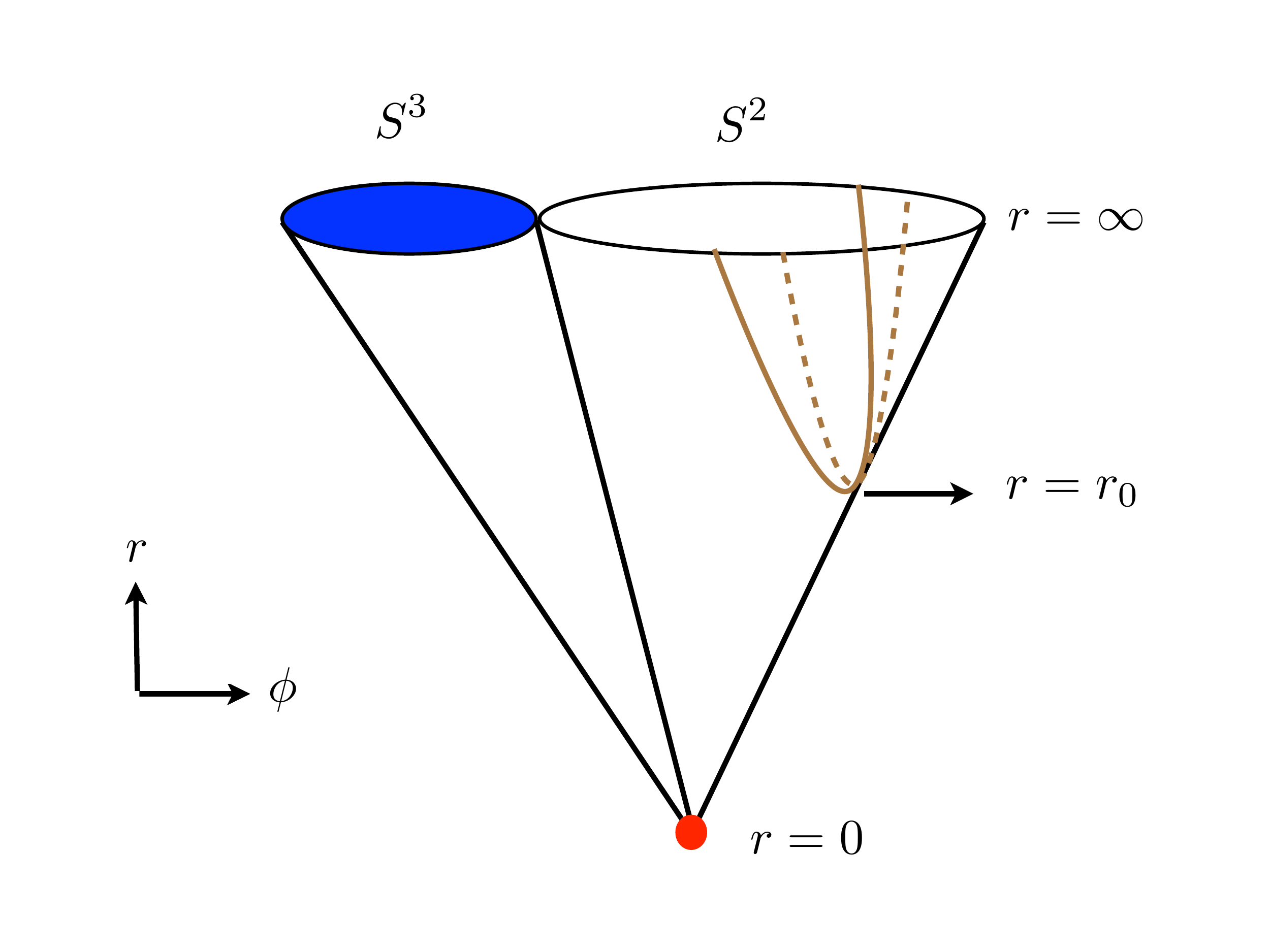}}
\caption{\small A schematic diagram showing the qualitative features of the brane--anti-brane profile in the presence of a non-zero magnetic field at zero temperature. The dashed grey curve represents the corresponding profile at zero magnetic field. The red dot again represents the conifold singularity located at $r=0$.}
\label{kwb}
\end{center}
\end{figure}

This equation of motion can be solved analytically and the solution is given by
\begin{eqnarray} \label{solh}
&& \cos \left(\frac{4}{\sqrt{6}} \phi(r) \right) = \left(\frac{r_0}{r}\right)^4 \frac{1}{h^2 + 2 r_0^4} \left[ h^2 \left(2 - \frac{r^4}{r_0^4} \right) + 2 r_0^4 \right] \ , \\
&& {\rm with} \quad \phi'(r_0) \to \infty \quad \implies \quad c = \frac{r_0^4}{\sqrt{6}} \left( 1 + \frac{h^2}{r_0^4} \right)^{1/2} \ ,
\end{eqnarray}
where $r_0$ is the point where the brane--anti-brane pair smoothly joins. In the limit $h \to 0$, we recover the known result in (\ref{sol0}). In fig.~\ref{kwb}, we have shown a schematic diagram of the shape of the probe brane profile.

A few comments on the asymptotic angle separation are in order. Considering the limit $r \to \infty$, from the solution in (\ref{solh}) we get
\begin{eqnarray} \label{delphi@h}
\Delta\phi_\infty = \frac{\sqrt{6}}{4} \pi + \frac{\sqrt{6}}{2} \alpha \ , \quad \alpha = \sin^{-1} \left( \frac{h^2}{h^2 + 2 r_0^4} \right) \ .
\end{eqnarray}
As a consequence of the explicit breaking of conformal invariance, the asymptotic angle separation is now promoted to a function of $h$ and $r_0$; in fact, it depends only on the dimensionless ratio $h/r_0^2$. It is clear from this expression that in the limit $h \to 0$, we recover the known result $\Delta \phi_\infty \to (\sqrt{6}/4) \pi$ and as $h\to \infty$, we get $\Delta \phi_\infty \to (\sqrt{6}/2) \pi$; for any intermediate value of $h$, $\Delta\phi_\infty$ interpolates between these two limiting values. The special solution for $r_0 = 0$ (which corresponds to $c =0$) is identified with the solution obtained at $h \to \infty$ limit and is simply given by: $\phi_{\pm} = \pm (\sqrt{6}/4) \pi$.

\begin{figure}[htp]
\begin{center}
 {\includegraphics[angle=0,
width=0.65\textwidth]{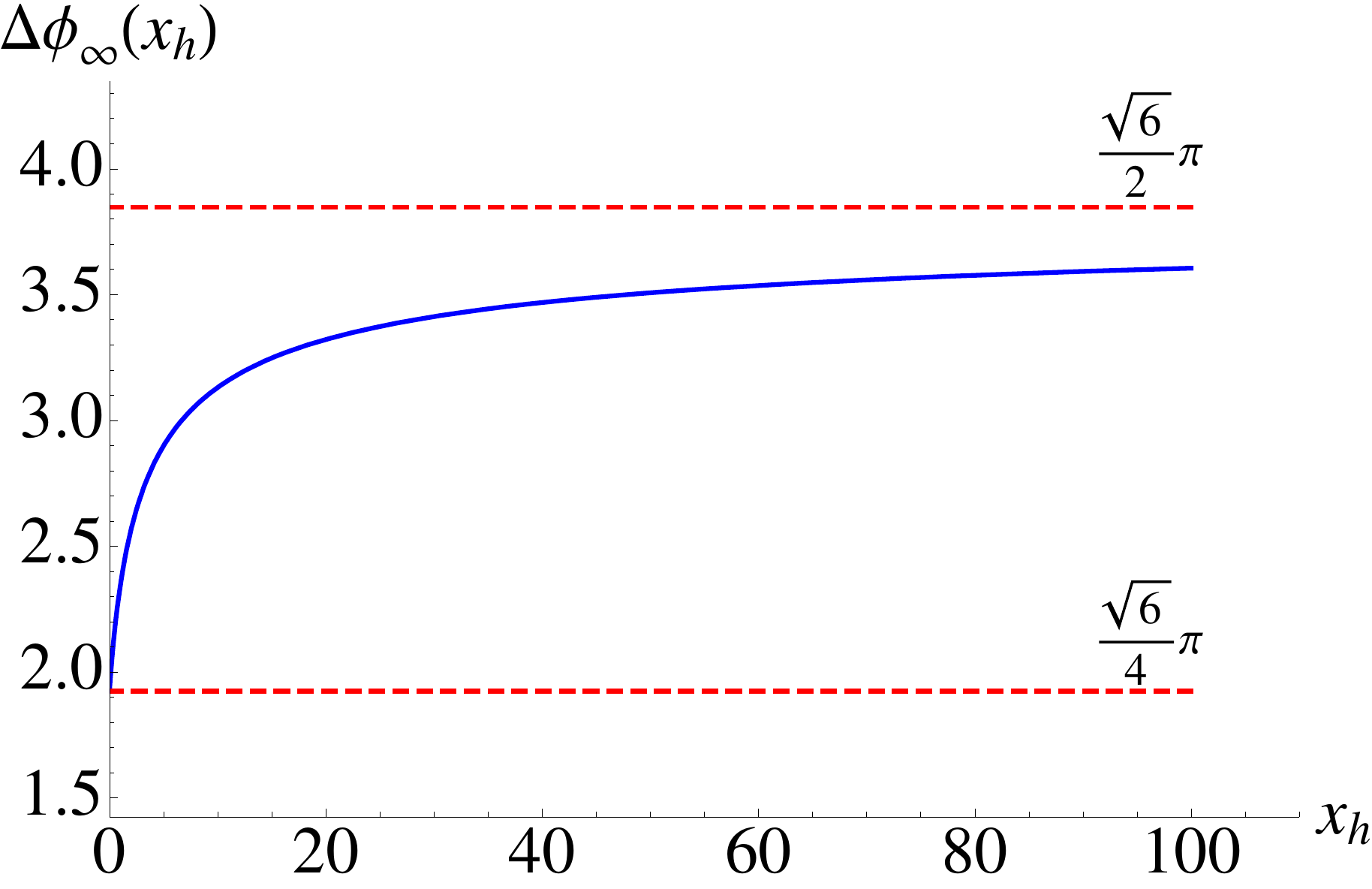}}
\caption{\small $\Delta \phi_\infty$ as a function of $x_h$. The range of allowed values are given by: $(\sqrt{6}/4) \pi \le \Delta\phi_\infty \le (\sqrt{6}/2) \pi$.}
\label{bc@x}
\end{center}
\end{figure}

For a more thorough investigation we obtain the following integral formula for the asymptotic angle separation
\begin{eqnarray}
\Delta \phi_{\infty} (x_h) & = & 2 \sqrt{6} \int_1^\infty \frac{dy}{y}  \frac{(1+x_h)^{1/2}}{\left[y^8 (1 + x_h/ t^4)  - (1+x_h)  \right]^{1/2}}  \nonumber\\
                                            & = & \sqrt{6} \tan^{-1} \left(\sqrt{1 + x_h} \right) \ , \\
                                          y & = & \frac{r}{r_0}\ , \quad x_h = \frac{h^2}{r_0^4} \ .
\end{eqnarray}
This is a monotonically increasing function of $x_h$. The dependence is explicitly demonstrated in Fig.~\ref{bc@x}.
From this monotonicity we can conclude that for a given $r_0$, $\Delta \phi_\infty(x_h) > \Delta \phi_\infty(0)$ which in turn implies that the magnetic field helps the brane--anti-brane pair to join. Since this is the basic mechanism leading to chiral symmetry breaking we expect that the magnetic field is further promoting this spontaneous symmetry breaking. We will find that this is indeed the case at finite temperature in the next section.

\section{Temperature and Magnetic Field}

Let us now consider the case where both temperature and magnetic field are present. For this we consider the background in (\ref{KW@T}) and we place the probe ${\rm D7}$/$\overline{{\rm D7}}$-brane similarly as before. The magnetic field is again realized as a gauge field on the worldvolume of the probe brane. The action is
\begin{eqnarray}  \label{lag@BT}
S & = & \tau_{7} \int d^8 \xi \sqrt{- {\det} \left (P [G + B] + (2 \pi \alpha' F \right)} = {\cal N}_T \int dr {\cal L} \nonumber\\
 & = & {\cal N}_T \int dr r^3 \left( 1 + \frac{h^2}{ r^4} \right)^{1/2} \left( 1 + \frac{r^2}{6} f(r) (\phi')^2 \right)^{1/2} \ .
\end{eqnarray}
The equation of motion is
\begin{eqnarray}
\frac{r^3 \left( 1 + h^2 /r^4 \right)^{1/2} (r^2/6) f \phi'}{\sqrt{1 + \frac{r^2}{6} f (\phi')^2}} = c \ .
\end{eqnarray}
This equation is not analytically solvable anymore. As in the pure finite temperature case, we have two different classes of solutions: the U-shaped ones and the $\parallel$ ones. The U-shaped ones are characterized by the position $r_0$ where the brane--anti-brane pair smoothly join which gives 
\begin{eqnarray}
c = \frac{r_0^4}{ \sqrt{6}} f(r_0)^{1/2} \left(1+ \frac{h^2} {r_0^4}\right)^{1/2} \ .
\end{eqnarray}
The $\parallel$ solutions are simply given by: $\phi_\pm(r) = \pm {\rm const}$ (which gives $c = 0$). These solutions corresponds to $0 \le \Delta\phi_\infty \le 2 \pi$.

On the other hand, for the U-shaped profiles the asymptotic angle separation is given by
\begin{eqnarray} \label{thxangle}
\Delta \phi_{\infty} (y_H, x_h) & = & 2 \sqrt{6} \int_1^\infty \frac{dy}{y} \sqrt{\frac{f(1)}{f(y)}} \frac{(1+x_h)^{1/2}}{\left[y^8 (1 + x_h/ y^4) f(y) - (1+x_h) f(1) \right]^{1/2}} \ , \\
x_h & = & \frac{h^2}{r_0^4} \ ,
\end{eqnarray}
where the ranges of the parameters are given by: $0 \le y_H \le 1$ and $0 \le x_h \le \infty$. In the limit $y_H \to 0_+$, $x_h \to 0_+$, we can analytically evaluate this integral to be given by
\begin{eqnarray} \label{delphi@TB1}
\Delta \phi_{\infty} (y_H, x_h) &  = & \frac{\sqrt{6}}{4} \left[ \left( \pi + x_h - \frac{1}{2} x_h^2 + {\cal O} (x_h^3)\right) \right. \nonumber\\
& +  & \left. \left( \frac{2 - \pi}{4} x_h + \frac{2\pi - 3}{8} x_h^2 + {\cal O} (x_h^3) \right) y_H^4 \right] + \ldots 
\end{eqnarray}
We can also evaluate this integral analytically in the limit $y_H \to 0_+$ and $x_h \to \infty$ to be given by the following
\begin{eqnarray} \label{delphi@TB2}
\Delta \phi_{\infty} (y_H, x_h) = \sqrt{\frac{3}{2}} \left[  \left( \pi - \frac{2} {\sqrt{x_h}} + {\cal O}(x_h^{-3/2})\right) + \left( - \frac{\pi}{4} + \frac{1}{\sqrt{x_h}} + {\cal O } (x_h^{-1})\right) y_H^4\right] + \ldots \ .
\end{eqnarray}
It is clear from both the expansions in (\ref{delphi@TB1}) and (\ref{delphi@TB2}) that $\Delta\phi_\infty $ approaches the respective constant values that we encountered in  sections \ref{T0} and \ref{TB}. In the limit $y_H \to 1$, however, it can be shown that this angle separation approaches zero as $\Delta \phi_\infty \sim (1 - y_H)^{1/2}$.

\begin{figure}[htp]
\begin{center}
\subfigure[] {\includegraphics[angle=0,
width=0.45\textwidth]{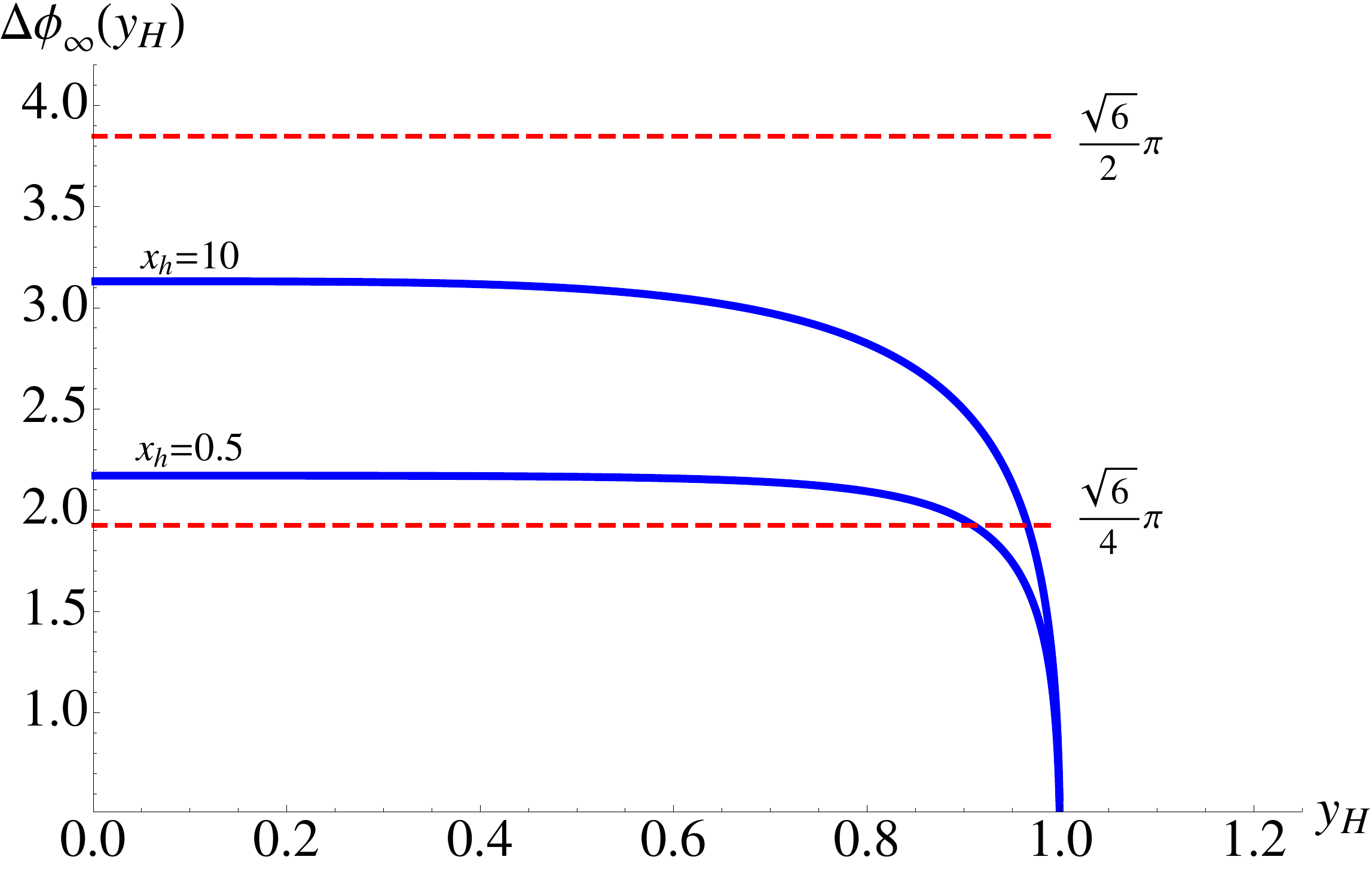} }
\subfigure[] {\includegraphics[angle=0,
width=0.45\textwidth]{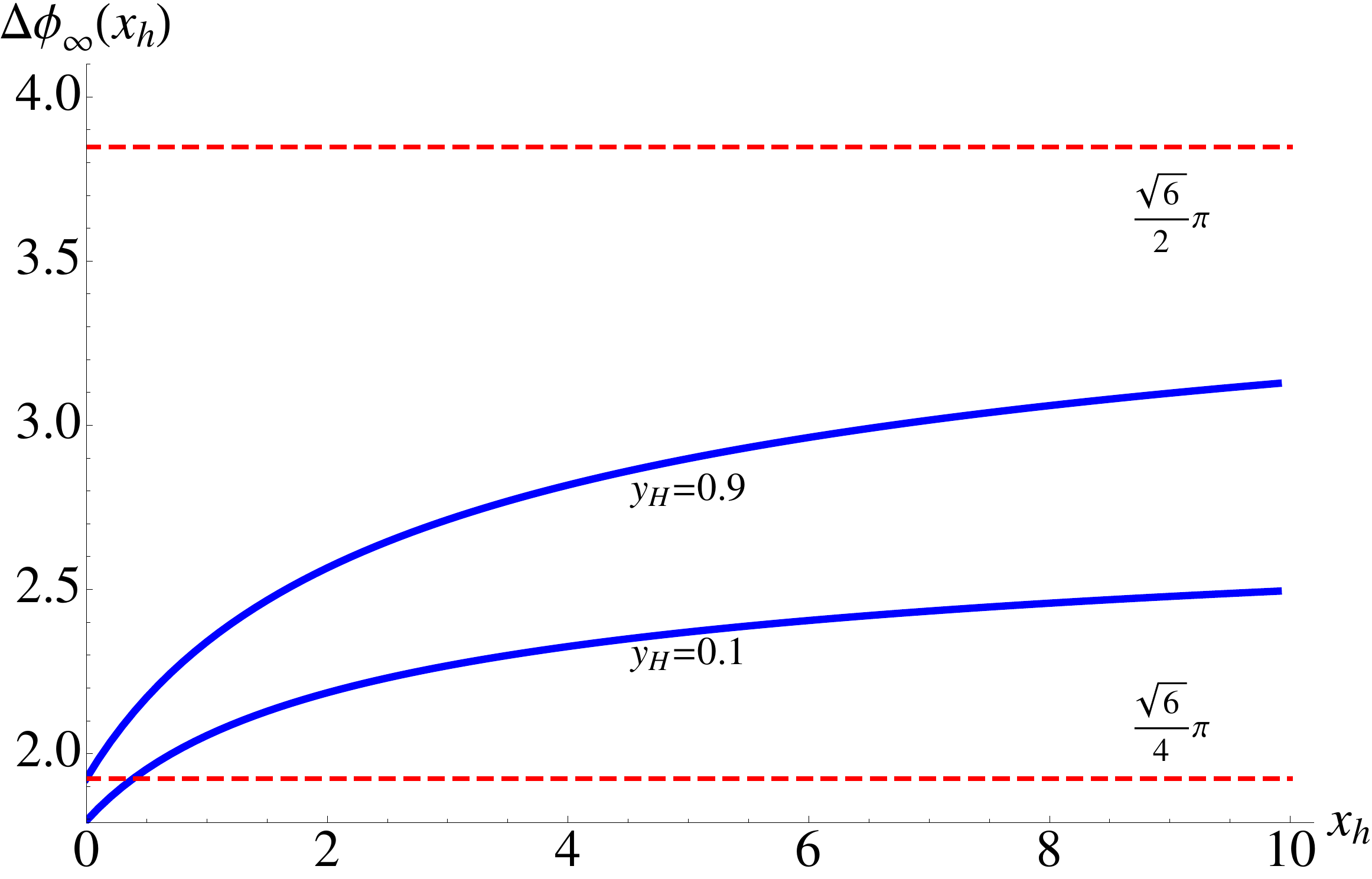} }
\caption{\small We have shown the dependence of $\Delta\phi_\infty$ as a function of $y_H$ (for a given value of $x_h$) on the left and as a function of $x_h$ (for a given value of $y_H$) on the right. It is clear from these plots that when we have both $y_H$ ad $x_h$ present, the allowed range is given by: $0 \le \Delta\phi_\infty \le (\sqrt{6}/2)\pi$ for the whole parameter space. Note that $\Delta\phi(y_H) \to 0$ as $y_H \to 1$, which is suggestive from the plot on the left.}
\label{bc-tH&x}
\end{center}
\end{figure}

For a generic point in the $\{y_H, x_h\}$ parameter space, we have to resort to numerics. Now $\Delta\phi_\infty$ depends on two variables $y_H$ and $x_h$ and generates a 3-dimensional plot, but we can take various constant $y_H$ or constant $x_h$-slices. Some such slices are shown in Fig.~\ref{bc-tH&x}.

\begin{figure}[htp]
\begin{center}
{\includegraphics[angle=0,
width=0.65\textwidth]{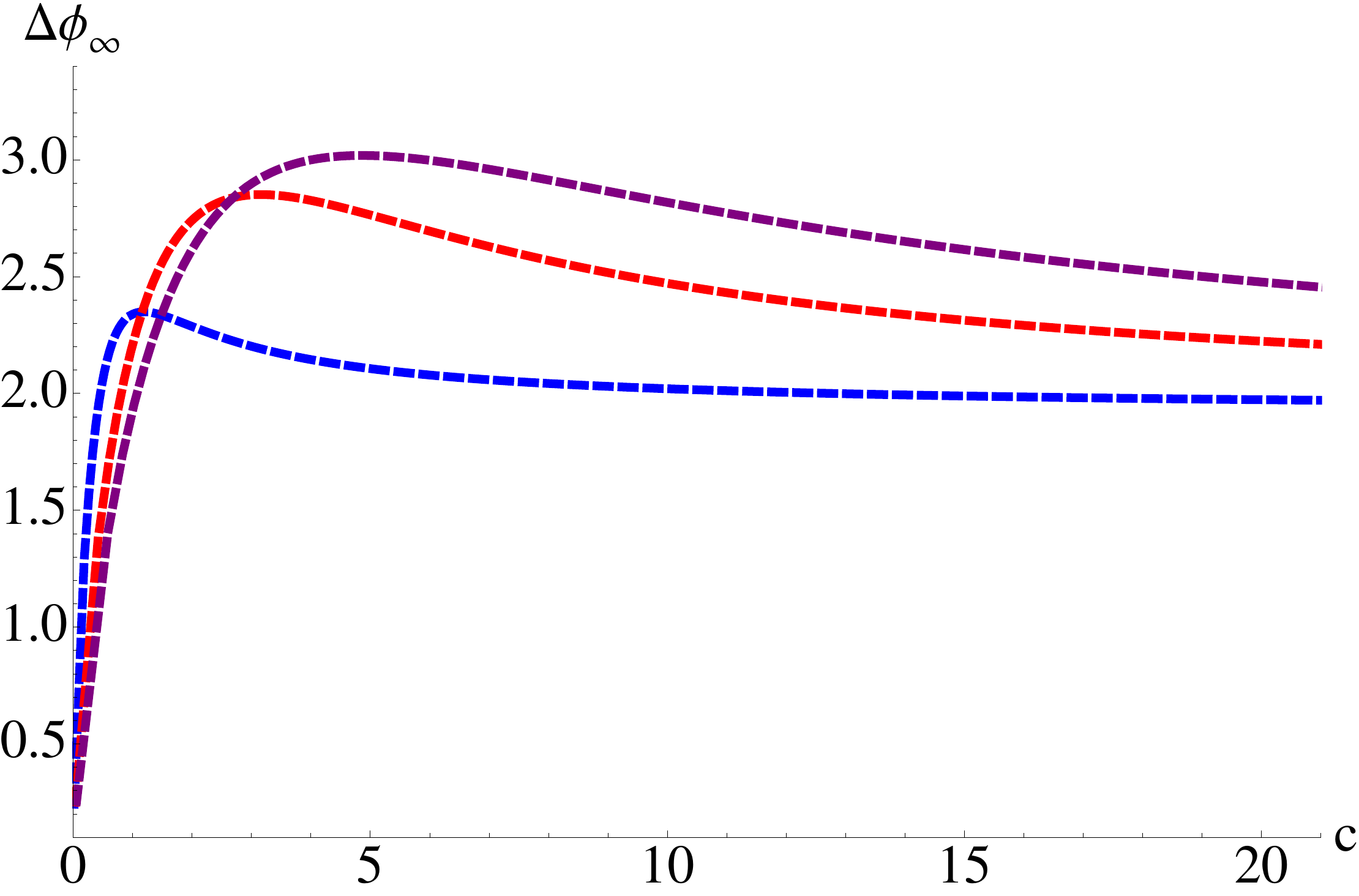}}
\caption{\small We have shown the dependence of $\Delta\phi_\infty$ as a function of $c$ for fixed values of $H/(\sqrt{\bar\lambda} T^2) = 2, 5, 7$ corresponding to blue, red and maroon curves. The asymptotic (as $c \to \infty$) value of $\Delta\phi_\infty$ for any value of $H/(\sqrt{\bar\lambda} T^2)$ approaches the constant value of $(\sqrt{6}/4) \pi$.}
\label{bending1}
\end{center}
\end{figure}

Before proceeding to determine the phase structure, let us analyze the possible phases closely. Since our underlying theory is conformal, the only meaningful quantity that we can vary is a dimensionless ratio constructed from the temperature and the magnetic field,
for example
\begin{equation}
\frac{h}{r_H^2} = \frac{H}{\sqrt{\bar\lambda} T^2}
\end{equation}
where $h = (2 \pi \alpha' R^2) H$ and $\bar\lambda=(\pi^2/4)\lambda_{\rm t\,Hooft}\,$;
we shall use this particular ratio in this and following sections.
Introducing a magnetic field (in the presence of a temperature) ultimately gives rise to the possibility of a first order phase transition. This can be best understood by looking at the $\{\Delta\phi_{\infty} - c \}$ plot as before. This is shown in Fig.~\ref{bending1}. The main qualitative difference as compared to the purely thermal case presented in Fig.~\ref{bc@c} is the bending of the curves for large enough values of $c$, which encodes the possibility of a first order phase transition.

\begin{figure}[htp]
\begin{center}
{\includegraphics[angle=0,
width=0.65\textwidth]{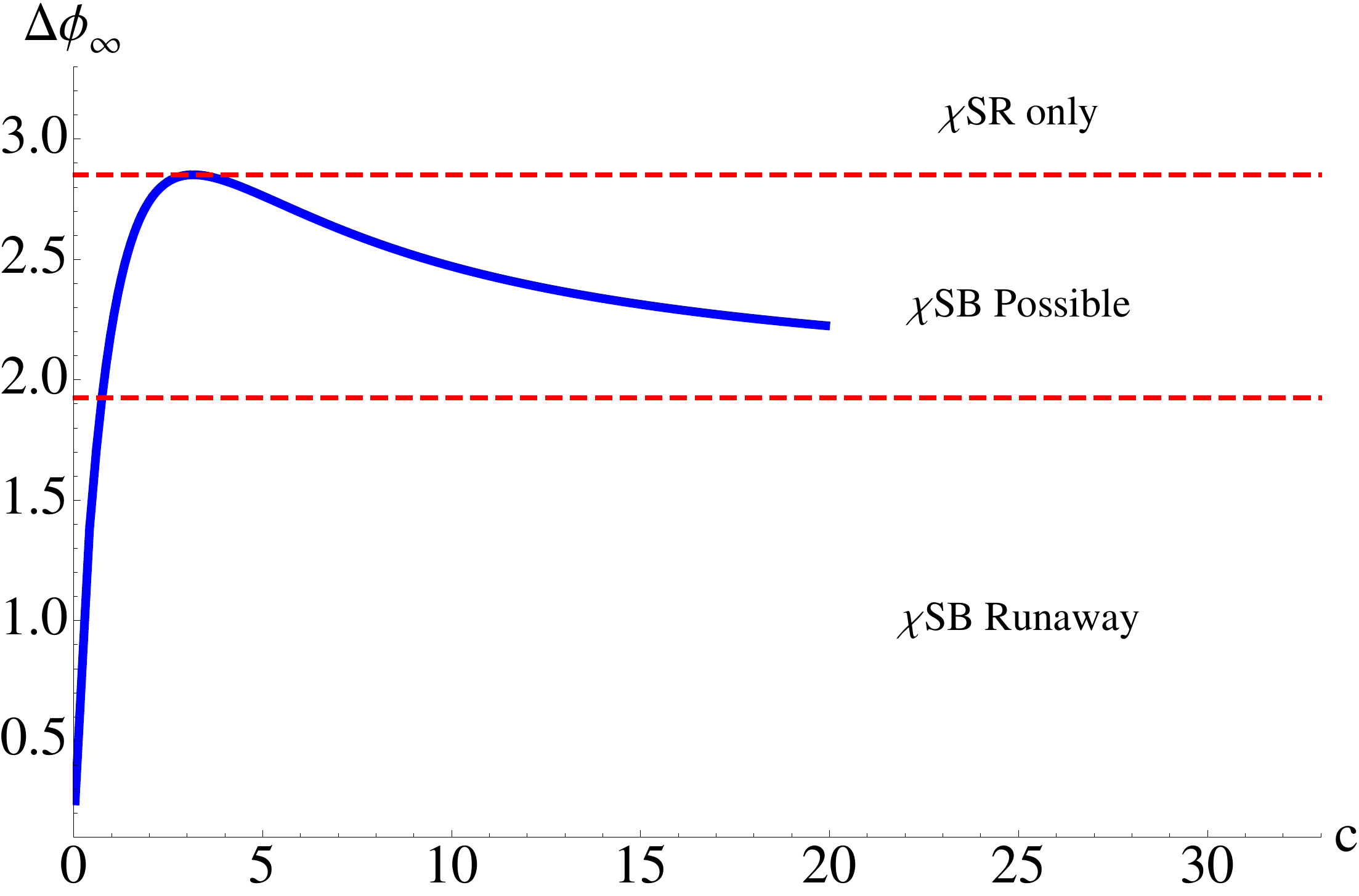}}
\caption{\small We have shown the dependence of $\Delta\phi_\infty$ as a function of $c$ for fixed values of $H/(\sqrt{\bar\lambda} T^2) = 5$. The asymptotic (as $c \to \infty$) value of $\Delta\phi_\infty$ for any value of $H/(\sqrt{\bar\lambda} T^2)$ approaches the constant value of $(\sqrt{6}/4) \pi$, which is marked by the lower horizontal red dashed line and the blue solid curve asymptotes to this line. We have marked the various possible phases for various ranges of the angle separation.}
\label{bending2}
\end{center}
\end{figure}

It is clear from Fig.~\ref{bending1} that the maximum value of $\Delta\phi_\infty$ depends on the value of $H/(\sqrt{\bar\lambda}T^2)$. For a given $H/(\sqrt{\bar\lambda} T^2)$, beyond the maximum value of $\Delta\phi_\infty$, there is no chiral symmetry broken phase. On the branch where $(\partial \Delta\phi_\infty) / (\partial c)>0$, increasing $r_0$ will increase the asymptotic angle separation and thus a small perturbation will either push the brane--anti-brane pair all the way up to infinity or pull them all the way down to the horizon. Thus the branch corresponding to $(\partial \Delta\phi_\infty) / (\partial c)>0$, although possesses the U-shaped embeddings,  is thermodynamically unstable. There is no chiral symmetry broken stable phase here. The only window where chiral symmetry broken phase can appear is for values of $\Delta\phi_\infty$ which lies in between its maximum value and the asymptotic value (as $c \to \infty$, demonstrated in Fig.~\ref{bending2}), where both the U-shaped and the parallel shaped embeddings are available and are thermodynamically stable. Within this window we need to compute the free energies of the corresponding phases to decide which embedding is thermodynamically favoured.

Now we check which embedding is picked by thermodynamic energy considerations within the window discussed above. To analyze what happens to the chiral symmetry, we need to evaluate the free energy difference which is given by (up to a factor of temperature) the difference of the on-shell Euclidean actions for the corresponding embedding 
\begin{eqnarray}\label{energydiff}
\Delta S & = & S_{U} - S_{\parallel} \nonumber\\
& = & {\cal N}_T r_0^4 \int_1^\infty dy y^3 \left( 1 + \frac{x_h}{y^4}\right)^{1/2} \left[ \left( 1 + \frac{(1+x_h) f(1)}{y^8 \left( 1 + x_h / y^4 \right) f(y) - (1 + x_h) f(1) }\right)^{1/2} - 1 \right] \nonumber\\
& - & {\cal N}_T r_0^4 \int_{y_H}^1
dy y^3 \left( 1 + \frac{x_h}{y^4}\right)^{1/2} =  {\cal N}_T r_0^4\, {\cal I} (y_H, x_h) \ .
\end{eqnarray}
We can argue that the right hand side of (\ref{energydiff}) changes sign for a given $y_H$ as we vary $x_h$. This can be seen from fixing the value of $y_H$ to be some very small non-zero number such that $y_H \ll 1$. Now in the limit $x_h \to 0_+$, we get $\Delta S \sim x_h \log y_H < 0$. On the other hand, in the limit $x_h \to \infty$ we get $\Delta S \sim \sqrt{x_h} y_H^2 >0$,  thus clearly indicating that $\Delta S$ goes through zero. Note that even for a small magnetic field $\Delta S$ starts off being negative which implies that the chiral symmetry broken phase is favoured. This broken symmetry now gets restored at some critical value of temperature. Thus our primary analysis indicates that the magnetic field is {\it catalyzing} in chiral symmetry breaking.

\begin{figure}[htp]
\begin{center}
{\includegraphics[angle=0,
width=0.95\textwidth]{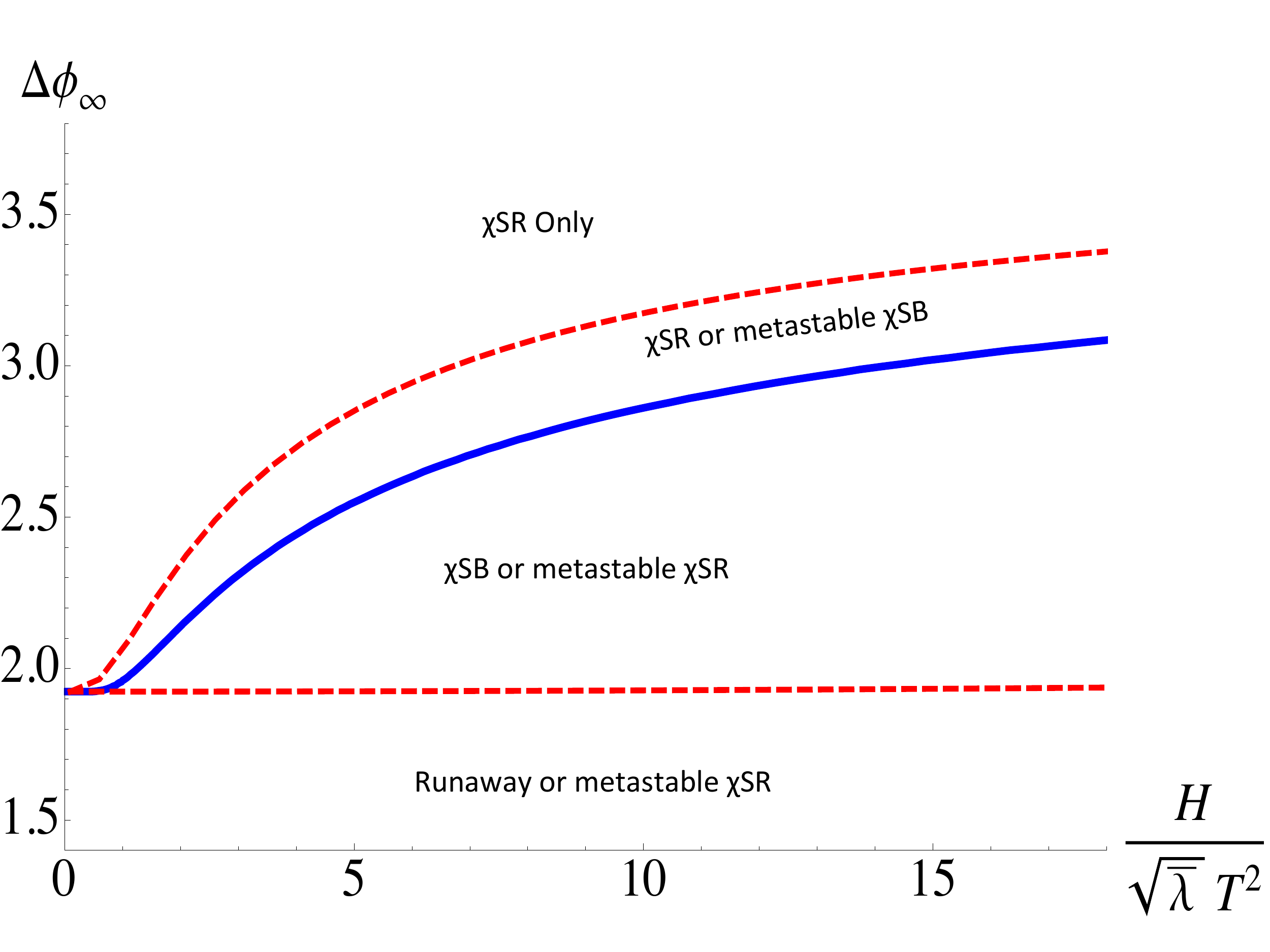} }
\caption{\small The phase diagram in the $\Delta\phi_\infty$ vs $H/ (\sqrt{\bar\lambda}T^2)$ plane. The upper red dashed curve corresponds to the maximum value of $\Delta\phi_\infty$ and the lower dashed red curve corresponds to the asymptotic (as $c\to \infty$) value of $\Delta\phi_\infty$. Below the red dashed line, we only have chiral symmetry restored phase for all values of $\Delta\phi_\infty$. We have not shown the complete range of this for aesthetic reasons.}
\label{TBphase}
\end{center}
\end{figure}

Finding the zeroes of the right hand side of (\ref{energydiff}) in the full parameter space will give a curve $y_H(x_h)$ which corresponds to the phase boundary between a chiral symmetry broken and a chiral symmetry restored phase. This phase boundary in the $\Delta\phi_\infty$ vs $H/(\sqrt{\bar\lambda} T^2)$-plane can be obtained numerically and the result is shown in Fig.~\ref{TBphase}.

In Fig.~\ref{TBphase} if we take the strict limit $H \to 0$, then there is no phase transition at all and all we have is the chiral symmetry restored phase for any given temperature. For a non-zero magnetic field the system undergoes a first order phase transition at some critical value of temperature for the range within the red dashed curves. For a fixed finite value of $H/(\sqrt{\bar\lambda} T^2)$, we get a critical $\Delta\phi_\infty$ below which the chiral symmetry broken phase is favoured and above which chiral symmetry is restored. As we increase the magnetic field, {\it i.e.} increase the ratio $H/(\sqrt{\bar\lambda} T^2)$, this critical coupling monotonically increases and in the strict $H \to \infty$ limit approaches the value $(\sqrt{6}/2) \pi$. Thus the external magnetic field indeed {\it catalyzes} the chiral symmetry breaking.

The appropriate thermodynamic potential for our system is the Helmholtz free energy given by 
\begin{eqnarray}
d {\cal F} = - {\cal S} dT - \mu d H \ , \quad {\cal F} = \frac{S}{T} \ ,
\end{eqnarray}
where ${\cal F}$ is the Helmholtz free energy, ${\cal S}$ is the entropy, $\mu$ is the magnetization and $S$ is the on-shell Euclidean action of the brane/anti-brane. The first order phase transition is associated with a non-zero latent heat and a relative change in magnetization given by
\begin{eqnarray}
{\cal S}_{\parallel} - {\cal S}_U & = & \Delta {\cal S}  =  -\left. \frac{\partial}{\partial T} \left( {\cal F}_{\parallel} - {\cal F}_{U} \right)\right |_{T_c} \ , \quad C_{\rm latent}  =  T_c \Delta {\cal S} \ . \nonumber\\
& = & \left. {\cal N}_T r_0^3 \pi R^2 \left( \frac{\partial {\cal I}}{\partial y_H} \right) \right |_{T_c}  \ , \\
\mu_{\parallel} - \mu_{U} & = & \Delta \mu  =  - \left. \frac{\partial}{\partial H} \left( {\cal F}_{\parallel} - {\cal F}_{U} \right)  \right |_{T_c} = \left. 2 {\cal N}_T r_0^2 R^2 \left( \sqrt{x_h} \frac{\partial {\cal I}}{\partial x_h}\right) \right |_{T_c}\ .
\end{eqnarray}
where ${\cal I}$ has been defined in (\ref{energydiff}). The absolute free energy is a formally divergent quantity, however the change in free energy is finite. The same is true for the magnetization. Thus instead of calculating the absolute quantities for each of these phases we focus on the relative ones.\footnote{To obtain the finite action for each phases, one needs to add proper counter terms to cancel the divergences. In this particular case, the free energy has two sources for divergences: one is a power law divergence which comes from the infinite volume of AdS and the other is a log-divergence  supported by a non-zero electromagnetic field strength.} 
\begin{figure}[htp]
\begin{center}
\subfigure[] {\includegraphics[angle=0,
width=0.45\textwidth]{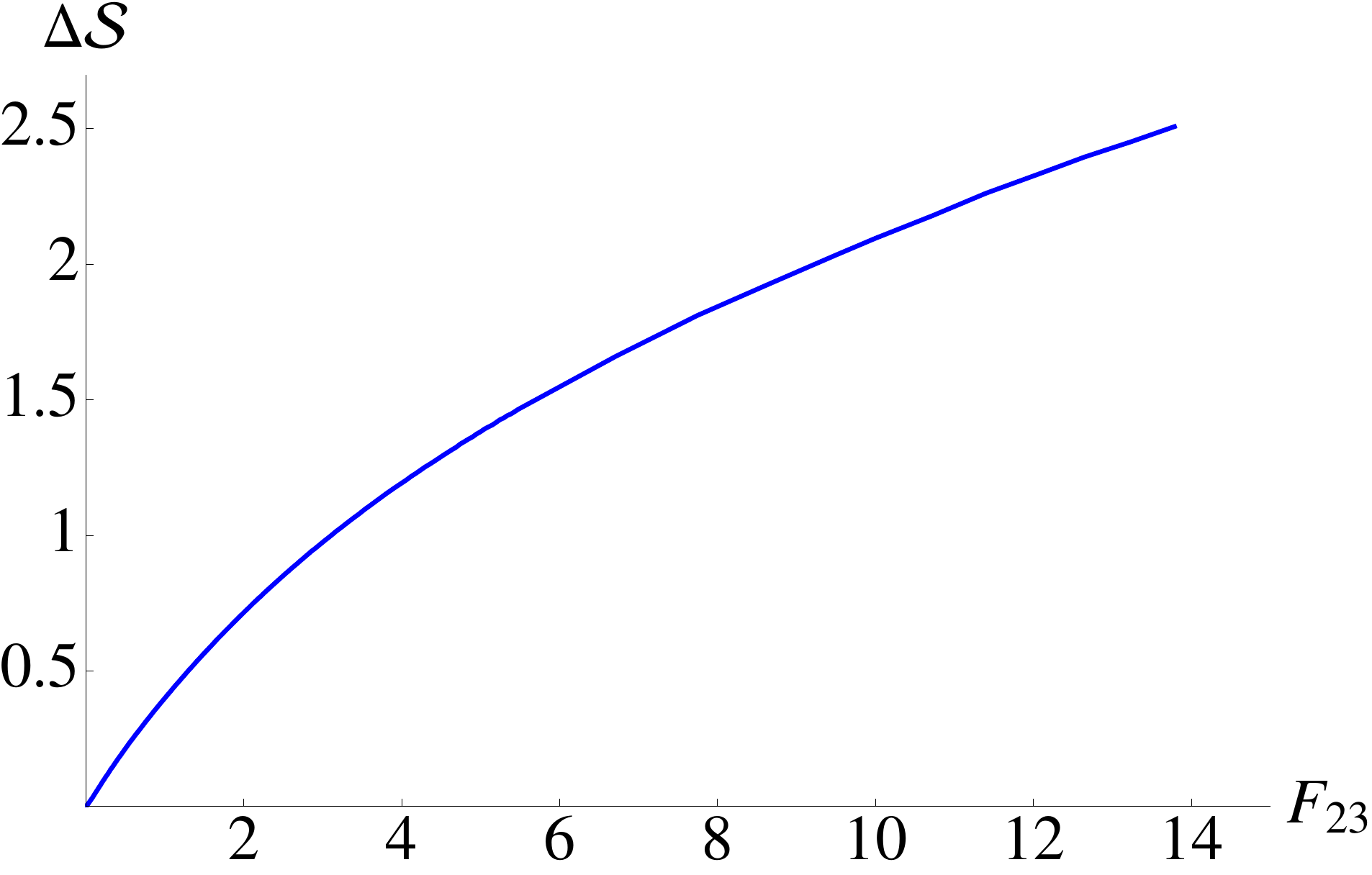} }
\subfigure[] {\includegraphics[angle=0,
width=0.45\textwidth]{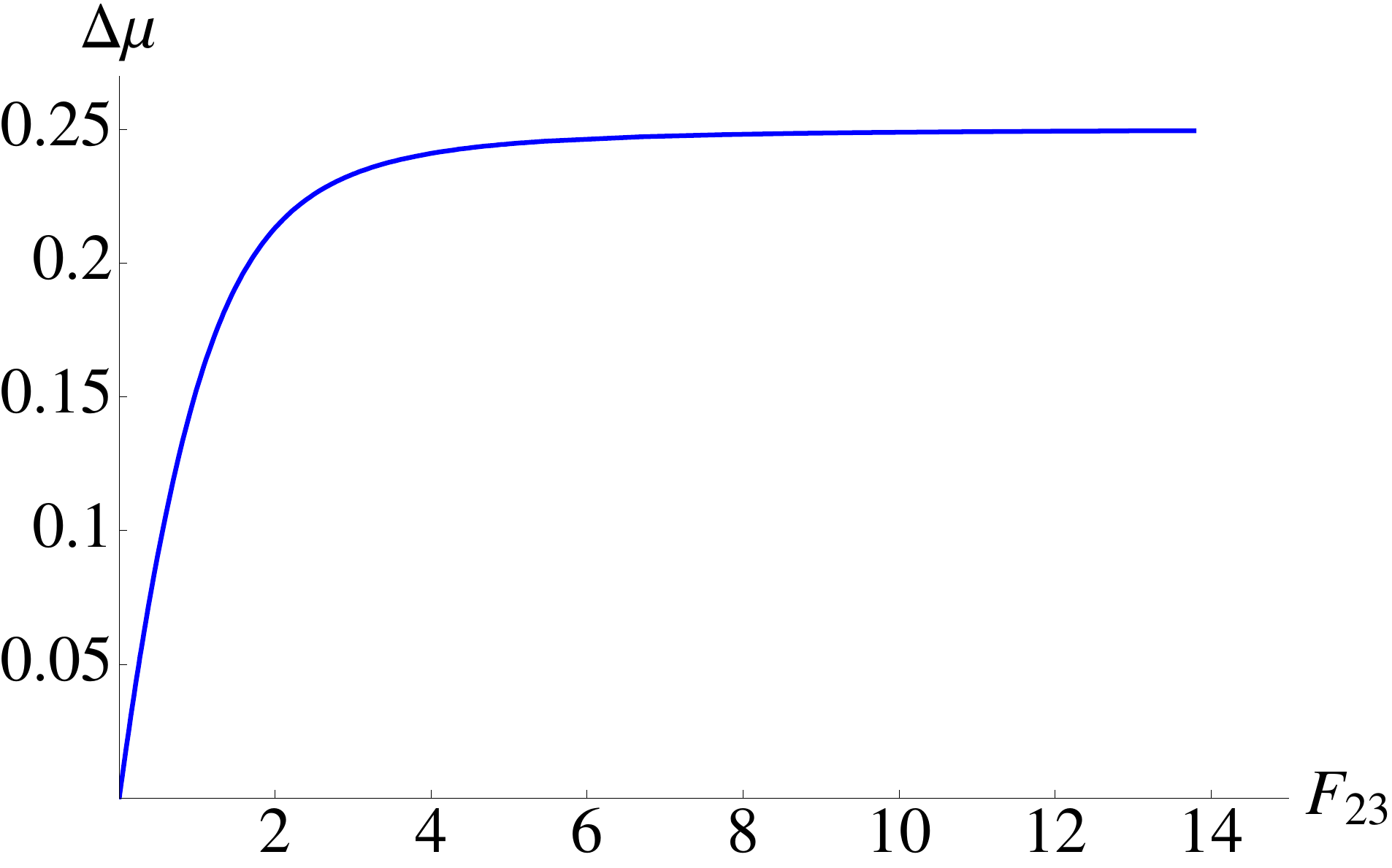} }
\caption{\small We have shown the dependence of the change in entropy and in magnetization associated with the first order phase transition. The entropy is measured in units of ${\cal N} r_0^3 \pi R^2 $, the magnetization is measured in units of $2 {\cal N} r_0^2 R^2$ and the magnetic field is measured in units of $ (2 \pi \alpha') (r_0/ R)^2$.}
\label{entrop-mag}
\end{center}
\end{figure}
 From simple scaling arguments it can be argued that both $C_{\rm latent} \sim T_c^4$ and $\Delta \mu \sim T_c^4$. Their dependence on the magnetic field is more involved and we have shown the numerical results in Fig.~\ref{entrop-mag}. Clearly the change in entropy increases with increasing magnetic field. The relative magnetization initially increases, but then seems to saturate an upper bound. The fact that $\Delta \mu > 0$ is intuitively clear: the chiral symmetry restored phase is more {\it ionized} than the chiral symmetry broken phase.

\begin{figure}[htp]
\begin{center}
{\includegraphics[angle=0,
width=0.65\textwidth]{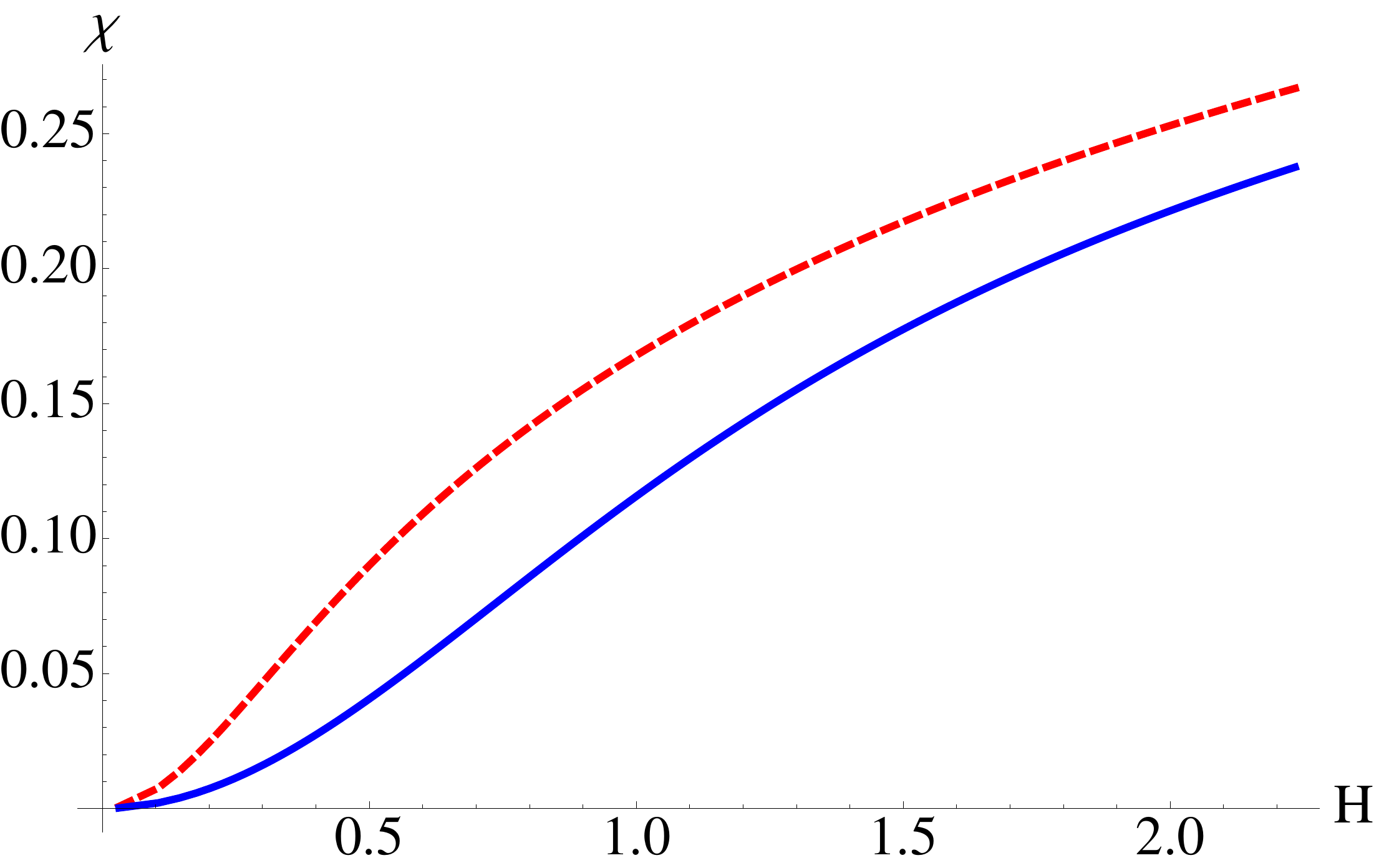}}
\caption{\small The dependence of the magnetic susceptibility with the applied magnetic field in the chiral symmetry broken (blue solid curve) and chiral symmetry restored (red dashed curve) phases. The magnetic susceptibility is measured in units of $(4 {\cal N} R^4)$ and the magnetic field is measured in units of $ (2 \pi \alpha') (r_0/ R)^2$.}
\label{chiUp}
\end{center}
\end{figure}

On the other hand, the log-divergence supported by the external field in the free energy is quadratic in the field strength. Thus any quantity obtained from taking the second derivative of the free energy with respect to the field strength can be regulated rather simply. One such thermodynamic quantity is the magnetic susceptibility. We define the regularized magnetic susceptibility as below\cite{Bergman:2008sg}
 \begin{eqnarray} \label{susreg}
 \chi = - \frac{\partial^2 {\cal F}}{\partial H^2} + \left. \frac{\partial^2 {\cal F}}{\partial H^2}  \right |_{H = 0 } \ .
 \end{eqnarray}
and compute the corresponding susceptibilities in each phases. For the chirally symmetric phase, the magnetic susceptibility (as defined in (\ref{susreg})) can be evaluated analytically to be given by 
\begin{eqnarray}
\chi_{\parallel} = - 2 {\cal N} R^4 \left[ 1 - \frac{y_H^2}{\sqrt{ y_H^4 + x_h }} - 2 \log (y_H)  + \log \left( \frac{y_H^2 + \sqrt{ y_H^4 + x_h}}{2} \right) \right] \ .
\end{eqnarray}
For the symmetry broken phase, the integral is not analytically tractable. The dependence of the magnetic susceptibility in both these phases has been shown in Fig.~\ref{chiUp}, which shows the non-linear monotonic dependence with the magnetic field. As expected, we also observe that the symmetry restored phase has higher susceptibility than the symmetry broken phase.
%

\section{Introducing an Electric Field}

We can include the effect of an external electric field in a similar manner. The gauge potential that we excite in this case is simply given by\cite{Karch:2007pd, Albash:2007bq}
\begin{eqnarray}
A_{x} = - E t + A (r) \ ,
\end{eqnarray}
where $E$ is the strength of the electric field along the $x^1$-direction.\footnote{Note that the ansatz for the gauge field contains the time coordinate $t$ explicitly. At the horizon (when a black hole is present in the background), this coordinate is ill-defined. A better coordinate system is the ingoing Eddington-Finkelstein coordinates. However, since the Poincar\'{e} coordinate is well-defined everywhere except at the horizon, it will not affect the physics we study here. For an analysis involving the Eddington-Finkelstein coordinates, see {\it e.g.} \cite{Karch:2010kt}.} Note that we have also included a function $A(r)$ in the ansatz for the gauge field. In the presence of an electric field we must also include the possibility of a flavour current. This function encodes this information of the current.

The Euclideanized\footnote{After Euclideanization, the gauge field will take the form: $A_x = - i E t_{\rm E} + A(r)$, where $t_{\rm E}$ is the Euclidean time obtained by $t \to i t_{\rm E}$.} DBI action takes the following form 
\begin{eqnarray} \label{actE}
S & = & {\cal N}_T \int dr r^3 \left[ \left( 1 - \frac{e^2}{f r^4} \right) \left( 1 + \frac{r^2}{6} f (\phi')^2\right) + f (a')^2 \right]^{1/2} \ , \nonumber\\
    & = & {\cal N}_T \int dr \cL(a', \phi', r) \ , \\
e & = & (2 \pi\alpha' E) R^2 \ , \quad a = 2 \pi \alpha' A \ .
\end{eqnarray}
From the structure of the action in (\ref{actE}) it is clear that for any non-trivial $\phi(r)$, the action is minimized for $a'=0$. The equation of motion for the profile function $\phi(r)$ is given by
\begin{eqnarray}
\left( 1 - \frac{e^2}{r^4 f(r) }\right)^{1/2} \frac{(r^5/6) f(r) \phi'}{\left( 1 + (r^2/6) f(r) \phi'^2\right)^{1/2}} = c = \frac{r_0^4}{\sqrt{6}} \left( 1 - \frac{e^2}{r_0^4 f(r_0) }\right)^{1/2} \ , 
\end{eqnarray}
where $r_0$ is the point where the brane--anti-brane pair joins. The reality condition of the constant $c$ imposes an upper bound for the electric field: $e^2 < r_0^4 f(r_0) = e_{\rm max}^2$. The existence of this maximum value of the electric field simply tells us that before we reach $e_{\rm max}$, the chiral symmetry restored phase should become energetically favourable. A similar effect has been discussed in \cite{Bergman:2008sg} for the Sakai-Sugimoto model. In our case, this is trivially true: at finite temperature the U-shaped embeddings are already energetically disfavoured. We expect an electric field will tend to restore the chiral symmetry, therefore our intuition tells us that at finite temperature and non-zero electric field, the parallel embeddings must be the thermodynamically preferred phase. We can check this explicitly by evaluating the corresponding on-shell actions; however, there are subtleties we need to address.\footnote{We are grateful to Oren Bergman and Gilad Lifschytz for a very fruitful discussion on this issue.}

Any non-trivial $a'(r)$ is supported only for the parallel embedding for which $\phi'=0$. For this class of embeddings, let us examine the action in (\ref{actE}) carefully. The variation of this action yields
\begin{eqnarray} \label{delS}
\delta S = {\cal N}_T \left [ \left. \frac{\partial \cL}{\partial a'} \delta a\right |_{r_{\rm min}}^{\infty} - {\rm EOM} \right ] \ ,
\end{eqnarray}
where $r_{\rm min}$ is the IR boundary. Usually this $r_{\rm min}$ should be identified with the location of the black hole horizon; however, we will argue this is not the case here. To have a well-defined variational problem the boundary term in (\ref{delS}) must vanish. We have the freedom to choose $\delta a(\infty) = 0$, but generically $\delta a(r_{\rm min}) \not = 0$. Thus in this case, we need to supplement the action in (\ref{actE}) with an additional boundary term at $r_{\rm min}$. This boundary term does not affect the equation of motion, but it will have a non-trivial effect when we evaluate the free energy by evaluating the on-shell action.

Before including this boundary term, let us look at the equation for the gauge field
\begin{eqnarray} \label{eomaphi}
&& \frac{\partial \cL}{\partial a'} = \frac{r^3 f a'}{\left[ \left( 1 - \frac{e^2}{f r^4} \right) + f (a')^2 \right]^{1/2}} = j \ , \\
&& a' = \pm \frac{j}{r^2 f} \sqrt{\frac{r^4 f - e^2}{r^6 f - j^2}} \ , \quad \implies \quad \left. a(r)\right |_{r\to \infty} = \mp \frac{j}{2 r^2} + \ldots \label{gauge}
\end{eqnarray}
where $j$ is a constant. It is clear that $a(r) \to 0$ as $r \to \infty$. We have to decide on the sign of the solution for $a'$ above. This can be fixed by imposing an ``ingoing" boundary condition at the horizon (meaning any energy-momentum flow at the horizon only flows into the horizon and not the other way round). This condition picks up the solution with the +ve sign.

Now let us go back to the action with the boundary term subtracted. This can be written as
\begin{eqnarray} \label{newactE}
S = \cN_T \left[ \int_{r_{\rm min}}^{\infty} dr \cL (a', r) - j \int_{r_{\rm min}}^{\infty} a' dr \right] \ ,
\end{eqnarray}
where we have used the fact that $a(r) \to 0$ as $r\to\infty$. From now on, in the presence of an electric field we will work with this action for the parallel embeddings. 

From the perspective of the boundary theory, the response current is given by
\begin{eqnarray}
\langle J_{x^1} \rangle = \lim_{\epsilon \to 0 } \left. \frac{1}{\sqrt{- \gamma}} \frac{1}{\epsilon^4} \frac{\delta S_{\rm ren}}{\delta A_{x^1}} \right |_{\epsilon} \ ,
\end{eqnarray}
where $\epsilon$ is the UV cut-off (in our notation $\epsilon \sim 1/r$) and $\gamma$ denotes the pull-back metric on the $r = 1/ \epsilon$ cut-off surface and $S_{\rm ren}$ is the renormalized action after adding the appropriate counter-terms. Following \cite{Karch:2007pd}, it can be easily shown that $\langle J_{x^1} \rangle \sim j$.

Let us now comment on the choice of $r_{\rm min}$ when we want to identify the on-shell action as the thermodynamic free energy. A natural choice is clearly $r_{\rm min} = r_H$. However there is an issue with this choice. The integral of $(j \cdot a')$ yields a contribution of the form $(j \cdot e) \tau$, where $\tau$ is some typical time-scale. This time-scale $\tau$ has an IR log-divergence coming from the horizon
\begin{eqnarray}
\tau = \frac{1}{4 r_H} \log\left( r - r_H\right) \quad {\rm with} \quad r \to r_H \ .
\end{eqnarray}
From the bulk point of view, $\tau$ can be identified with the time light rays take to travel from the boundary (at $r=\infty$) to the horizon (at $r=r_H$)\cite{Karch:2008uy}. On the other hand, from the gauge theory point of view, the existence of $(j \cdot e) \tau$ can be interpreted as the total energy dissipated to maintain the current $j$ from time $t=0$ to time $t=\tau$.

The process of switching on an electric field and the onset of flow of charges is a time-dependent one. The electric field creates fundamental matter anti-matter pair via the Schwinger mechanism and accelerates them. This results in a deposition of energy into the background thermal bath of the adjoint matter. In the probe limit, the energy density of the fundamental sector is suppressed by a factor of $N_f/N_c$ and thus the background does not heat up. Finally the fundamental matter reaches a steady-state where a constant current flows.

Thus what we have here is not a stationary equilibrium state. Clearly, the energy dissipated to maintain the current should not be included in the thermodynamic free energy of the corresponding phase. The physics is telling us the choice of $r_{\rm min} = r_H$ is incorrect as far as the computation of the thermodynamics goes and there has to be another radial scale naturally arising in this problem. This is indeed the case.

The parallel branes go all the way to the horizon. Thus the on-shell action in this case is given by
\begin{eqnarray}
S = {\cal N}_T \left[ \int_{r_H}^{\infty} dr \left\{ r^3 \left( 1 - \frac{e^2}{r^4 f(r)} \right)^{1/2}  \left( 1  - \frac{j^2}{r^6 f(r)} \right)^{-1/2} -  j a' \right\}\right] \ .
\end{eqnarray}
This action must remain real. This reality condition imposes two algebraic conditions. These conditions determine the constant $j$ in terms of $e$ and $r_H$
\begin{equation}
e^2\,=\,r_*^4 f(r_*)\quad{\rm and}\quad j^2\,=\,r_*^6\,f(r_*)\quad
{\rm for\ the\ same}\ r_*\ ,
\end{equation}
hence
\begin{equation}
r_*^4\,=\,r_H^4\,+\,e^2
\end{equation}
and
\begin{equation} \label{currentKW}
j\,=\,e\times\root 4 \of{r_H^4+e^2}\ .
\end{equation}
Thus we obtain the analogue of an Ohm's law where the conductivity depends non-linearly on the electric field. This result is in precise agreement with the one obtained in \cite{Karch:2007pd, Albash:2007bq} by considering a completely different kind of D7-brane embedding in AdS-Schwarzschild$\times S^5$ background.\footnote{This corresponds to adding ${\cal N} = 2$ hypermultiplets to the ${\cal N}=4$ super Yang-Mills.} This conductivity depends entirely on the non-compact part of the background metric (which is $AdS_5$ here) and is insensitive to the details of how the probe brane is embedded along the compact internal directions. Although the dual gauge theories are different in these cases, this fact tells us that the finite temperature transport properties (such as the conductivity) are insensitive to such differences.

The above algebraic constraints do more for us than to just determine the conductivity; they give us another natural radial scale denoted by $r_*$. Following \cite{Albash:2007bq}, we will call this the ``pseudo-horizon". We will argue momentarily that this radial scale acts as a natural ``cut-off" as far as thermal properties are concerned, which is otherwise usually played by the event-horizon. Note that at this radial position nothing special happens to the background; moreover the induced metric on the probe D7-branes is also ignorant about the location of the pseudo-horizon. As we will argue now, it is only the flavour degrees of freedom which are sensitive to the existence of the pseudo-horizon.

Before doing so, let us remind ourselves some important facts on the physics at finite temperature. At finite temperature, we identify the Euclidean on-shell action of the probe with its thermodynamic free energy (up to a factor of temperature). In evaluating the Euclidean on-shell action, we use $r_{\rm min} = r_H$. As elaborated in \cite{Karch:2008uy} with a toy model, the action of the probe that is inside the black hole contributes to the overall entropy of the background once the back-reaction of the probe is taken into account. We do not need to account for the part of the probe D-brane inside the horizon while computing the thermodynamic free energy of the probe sector.

Now, the degrees of freedom living on the probe brane are the open string degrees of freedom. In the presence of a background gauge field, {\it e.g.} an electric field on the probe world-volume, the effective geometry perceived by the fundamental sector can be different from the background geometry. As has been explicitly demonstrated in the seminal work in \cite{Seiberg:1999vs}, in the presence of background gauge fields, the open string ``feels" an effective geometry described by the so called open string metric. Let $G$ be the induced metric on the probe and $F$ be the constant electromagnetic field on its world-volume, then the open string metric, denoted by $\cS$, is given by
\begin{eqnarray}
\cS_{ab} = G_{ab} - \left(F G^{-1} F\right)_{ab} \ .
\end{eqnarray}
This metric can be seen to naturally arise by expanding the DBI Lagrangian to quadratic order. In the presence of an electric field it can be explicitly shown that this open string metric $\cS$ has a horizon at $r=r_*$. We have explicitly demonstrated this in Appendix C. In \cite{Kim:2011qh}, using a similar set-up (studying D7-branes in AdS-Schwarzschild$\times S^5$-background) it has also been explicitly shown that the various conductivities can be determined by the data at $r = r_*$. This is reminiscent of the ``membrane paradigm"; however the ``fictitious" membrane is not located at the horizon, but at the pseudo-horizon.

Thus, as far as the fundamental degrees of freedom are considered, the pseudo-horizon plays an analogue role of the actual event-horizon of the space-time. Also, from the analysis of the different classes of embeddings in the presence of an electric field, we conclude that once the probe brane crosses $r=r_*$, it has to turn on a current $j$ given by the formula in (\ref{currentKW}) and fall all the way through the horizon. Therefore, we propose to identify the Euclidean on-shell action evaluated up to $r=r_*$ with the thermodynamic free energy of the probe in the presence of an electric field. Notice, however, the pseudo-horizon has an important difference compared to an event-horizon: classically we cannot recover any information hidden behind an event-horizon; whereas information can propagate outside the pseudo-horizon. In analogy with the analysis done in \cite{Karch:2008uy}, we conjecture that the part of the probe brane hidden behind $r=r_*$ contributes to the production of entropy of the background once the back-reaction of the probes are taken into account. It will be extremely interesting to verify this claim explicitly, but this is a non-trivial problem which we leave for future investigations.

Now, let us comment on a technical advantage of using $r_{\rm min} = r_*$. Recall that choosing $r_{\rm min} = r_H$ led to an IR log-divergence. It can now be explicitly checked that the action in (\ref{newactE}) is perfectly IR-finite if we choose $r_{\rm min} = r_*$. Emboldened by all these observations, we propose the following prescription for computing the free energy in the presence of the electric field. We compute the on-shell action, but truncate it in the IR at $r=r_*$. Note that in determining $r_*$ and $j$ we can either use the on-shell action extended all the way to the horizon or we can simply impose reality condition for the solution of the gauge field in (\ref{gauge}). The boundary term that we added in (\ref{newactE}) can be interpreted as follows: for the part of the brane above $r_*$, this term simply acts as a boundary term; for the part of the brane below $r_*$, this acts as a source.

Now, with our conjectured proposal, we can easily verify that the parallel embeddings are always energetically favourable and chiral symmetry is always restored for the purely electric field case. We will get non-trivial phase structure in the presence of both electric and magnetic field at finite temperature, which we study in the next section.

\section{Electric and Magnetic Field}

We have argued and explicitly shown that an external magnetic field helps in chiral symmetry breaking whereas an external electric field restores the symmetry. Clearly electric and magnetic fields are two competing parameters as far as chiral symmetry breaking is considered. In this section we will explore the corresponding phase diagram when both of these competing parameters are present at finite temperature.

So far the dynamics of the flavours have been governed solely by the DBI action. In the presence of the electric and magnetic field (specifically the case when they are parallel to each other as we will see later), there will be a non-zero contribution coming from the Wess-Zumino term as well. This term takes the following general form
\begin{eqnarray}
S_{\rm WZ} = \mu_7 \int \sum_p C_p \wedge e^{2\pi \alpha' F + B} \ ,
\end{eqnarray}
where $\mu_7$ is related to the 7-brane tension, $F$ is the worldvolume 2-form field strength, $B$ is the NS-NS 2-form and $C_p$ is the $p$-form potential present in the background. The supergravity background given in (\ref{kw}) does not have any NS-NS field and $F_5$ is the only Ramond-Ramond field strength that is present. Thus the non-zero contribution coming from the Wess-Zumino term in this case takes the following general form
\begin{eqnarray} \label{cs}
S_{\rm WZ} = \frac{\mu_7}{2} \int P[C_4] \wedge F \wedge F +  \frac{\mu_7}{2} \int P[\tilde{C}_4] \wedge F \wedge F \ ,
\end{eqnarray}
where $P$ denotes the pull-back and the potentials $C_4$ and $\tilde{C}_4$ are defined by
\begin{eqnarray}
&& F_5 = d C_4 \ , \\
&& \star F_5 = d \tilde{C}_4 \ . 
\end{eqnarray}
Here $\star$ represents the $10$-dimensional Hodge dual. The explicit form of the potentials are given by
\begin{eqnarray}
&& C_4 = \frac{1}{g_s} \frac{r^4}{R^4} dt \wedge dx \wedge dy \wedge dz \ , \\
&& \tilde{C}_4 = - \frac{R^4}{27 g_s} \cos\theta f_1 \wedge f_2 \wedge f_3 \wedge d\phi \ .
\end{eqnarray}
It is clear from the expression in (\ref{cs}), since $F$ has legs along the Minkowski directions only, the Wess-Zumino term (which is proportional to $F \wedge F$) gives a non-zero contribution when the Minkowski electric and magnetic fields are parallel and this contribution comes solely from the second term in (\ref{cs}). We will discuss the consequences of this term in a subsequent subsection.

At zero temperature, there are only two Lorentz invariants: $\vec{E}^2 - \vec{H}^2$ and $\vec{E} \cdot \vec{H}$. Thus it suffices to consider two configurations: $ \vec{E} \perp \vec{H} $ and $\vec{E} || \vec{H} $. Non-zero finite temperature breaks this Lorentz invariance. For a generic configuration, both the DBI and the WZ contributions depend on the relative angle of the electric and the magnetic fields. For simplicity, here we will focus on two representative cases: $\vec{E} \perp \vec{H} $ and $\vec{E} || \vec{H} $. In view of our discussion earlier, the WZ piece will contribute only in the parallel configuration. In this process, we will obtain the corresponding formulae for the flavour conductivity for these two cases. As before, these formulae are identical to the ones obtained in \cite{O'Bannon:2007in}.

\subsection{The case of perpendicular fields}

Let us first consider the case when the electric and the magnetic fields are perpendicular
and the Chern--Sinons term vanishes.
Our ansatz for the gauge fields is\footnote{Note that in general one would expect the presence of a Hall current perpendicular to the electric field for this configuration; however, it can be shown explicitly (or see {\it e.g.} \cite{O'Bannon:2007in}) that this Hall current is proportional to the chemical potential in this system. We do not consider the theory at finite chemical potential, hence we are safe to ignore the Hall current.}
\begin{eqnarray}
A_x = - E t + A(r) \ , \quad A_y = H x \ .
\end{eqnarray}
and the probe action is  given by
\begin{equation}
\label{perpaction}
S\ =\ {\cal N}_T \left[
	\int dr r^3 \left[ \left( 1 + \frac{h^2}{r^4} - \frac{e^2}{f r^4} \right)
		\left( 1 + \frac{r^2}{6} f (\phi')^2\right) + f (a')^2 \right]^{1/2}
	- j \int dr a'
	\right] ,
\end{equation}
where
\begin{equation}
e\ =\ (2 \pi\alpha' E) R^2, \quad h = (2 \pi\alpha' H) R^2 ,\quad
a'(r)\ =\ 2 \pi \alpha'\frac{dA}{dr}\,.
\end{equation}
Note that $e$ and $h$ have dimensions $\rm length^2$, so the phase structure of the theory
depends on the dimensionless ratios $e/h=E/H$ and
\begin{equation}
\frac{h}{r_H^2}\ =\ \frac{H} {\sqrt{\bar \lambda} T^2}
\end{equation}
where $\bar\lambda=(\pi^2/4)\lambda_{\rm t\,Hooft}$.

The first term on the action (\ref{perpaction}) is the DBI action while the second
term is a total derivative  which  does not change the equations of motion
but contributes to the net action of the brane.
As we saw in section~6, for the U-shaped brane embedding the boundary
conditions at the two sides of the U lead to $a(r)=\rm const$, but
for the $||$ embedding there is $a'(r)\neq0$.
Solving the equation of motion for the $a(r)$, we obtain
\begin{equation}
a'(r)\ =\ \frac{j}{r^2 f} \left(
	\frac{\left(r^4 + h^2 \right) f - e^2 }{r^6 f - j^2 }\right)^{1/2}
\end{equation}
where $j$ is the constant determined from the reality of the action integral
\begin{equation}
S_{||}\ =\int^\infty_{r_{\rm min}}\cL_{||}dr\
=\ {\cal N}_T  \int^\infty_{r_{\rm min}}\!dr \left[
	r^4\left( \frac{\left(h^2 + r^4\right) f(r) - e^2}{r^6 f(r) - j^2 }\right)^{1/2}\,
	-\,ja'(r)\right] .
\end{equation}
As explained in section~6, the lowest poin $r_{\rm min}$ of this integral is
the pseudohorizon $r_*$.
For reality's sake, both the numerator and the denominator of the ratio under
the square root must change signs at the same point $r=r_*$, hence
\begin{eqnarray}
&& \left(h^2 + r^4\right) f(r) - e^2 = 0 \quad \implies \quad r_* ^4 = \frac{1}{2} \left( \left(e^2 + r_H^4 - h^2\right) + \sqrt{\left(e^2 + r_H^4 - h^2\right)^2 + 4 h^2 r_H^4}\right) \ , \nonumber\\
&& j^2 = r_*^6 f(r_*) \ .
\end{eqnarray}
The above result matches with \cite{O'Bannon:2007in} in the appropriate limit.

For the U-shaped embeddings $a'\equiv0$ while the equation of motion for the $\phi'(r)$
is
\begin{equation} \label{phiebperp}
\frac{(r^5/6) \left( 1 + \frac{h^2}{r^4} - \frac{e^2}{f r^4} \right)^{1/2} f \phi'}%
	{\left( 1 + \frac{r^2}{6} f (\phi')^2 \right)^{1/2}}\
=\ {\rm const}\
=\ \frac{r_0^4}{\sqrt{6}} \sqrt{f(r_0)}
	\left( 1 +  \frac{h^2}{r_0^4} - \frac{e^2}{f(r_0) r_0^4}\right)^{1/2} .
\end{equation}
The on-shell action for this class of embeddings becomes
\begin{eqnarray} \label{suebperp}
S_U & = & {\cal N}_T  \int_{r_0}^{\infty} dr \frac{r^3}{\sqrt{f(r)}} \left( (r^4 + h^2) f(r) - e^2 \right) \left[ \frac{1}{r^4 \left((r^4 + h^2) f(r) - e^2 \right) - \left( (1 + h^2) f(1) - e^2 \right)} \right]^{1/2}  \nonumber\\
         & = & \int_{r_0}^{\infty} \cL_{U} dr \ .
\end{eqnarray}

Now we have to resort to numerical analysis to find out the thermodynamically preferred embedding. According to our proposal for the free energy, the corresponding phase diagram is obtained by looking at the zeroes of the following energy difference
\begin{eqnarray}
\Delta S & = &  \int_{r_0}^{\infty} \cL_{U} dr -  \int_{r_*}^{\infty} \cL_{||} dr \ .
\end{eqnarray}
Notice that our underlying theory was conformal. We have introduced three dimensionful scales in the system (the temperature, the electric and magnetic field), which explicitly break the conformal invariance. However, the only meaningful quantities we can talk about are two dimensionless ratios: $E/H$ and $H/(\sqrt{\bar\lambda}T^2)$. Thus our goal will be to study the dependence of $\Delta\phi_{\infty}$ as a function of each of these ratios for a fixed value of the other one.

\begin{figure}[htp]
\begin{center}
{\includegraphics[angle=0,
width=0.65\textwidth]{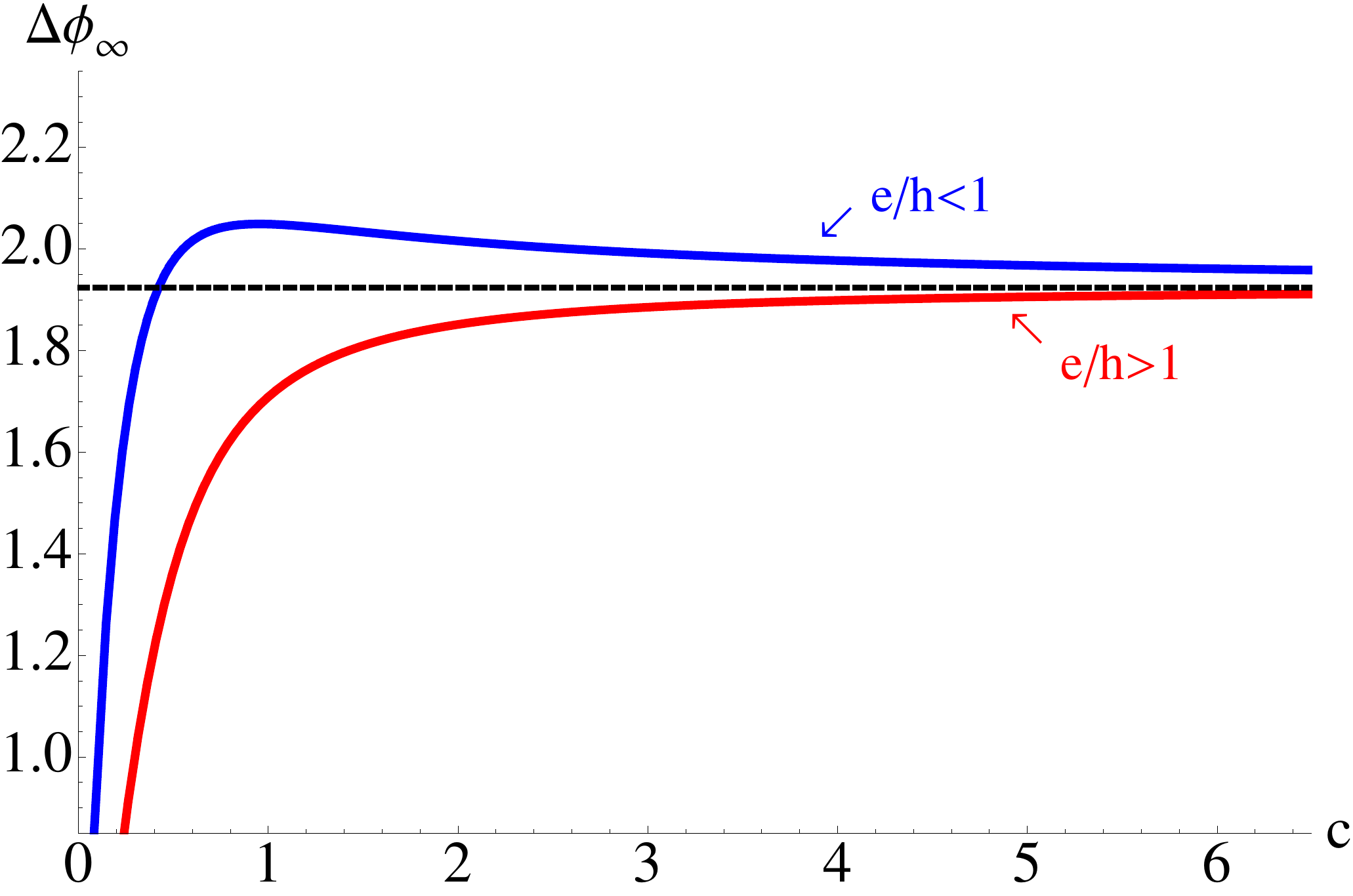}}
\caption{\small We have shown the dependence of the asymptotic angle separation as a function of $c$ for both $e/h>1$ and $e/h<1$. It is clear that for $e/h>1$, there is no phase transition and the interesting physics can happen only in the regime $e/h<1$. The black (horizontal) dashed line represents the value $\sqrt{6} \pi/4$.}
\label{pebperp}
\end{center}
\end{figure}

Before proceeding further, let us investigate some important features of the asymptotic angle separation in this case. From equation (\ref{phiebperp}), the asymptotic angle separation is given by
\begin{eqnarray}
\Delta\phi_\infty = \frac{3c}{x_0} \int_1^{\infty} \frac{dy}{\sqrt{y (y-1)}} \frac{1}{\sqrt{y - \frac{r_H^4}{x_0}}} \frac{1}{\sqrt{ y + 1 + \frac{\Delta}{x_0} }} \ ,
\end{eqnarray}
where we have defined
\begin{eqnarray}
x_0 = r_0^4 \ , \quad \Delta = h^2 - e^2 - r_H^4 \ , \quad y = \frac{r^4}{r_0^4} \ ,
\end{eqnarray}
and
\begin{eqnarray}
3 c = \sqrt{\frac{3}{2}} \left( x_0^2 + x_0 \Delta - r_H^4 h^2 \right)^{1/2} \ .
\end{eqnarray}
In the limit of large $c$, which translates to the limit of large $x_0$, we can obtain the following formula:\footnote{Interestingly, there is a term proportional to $r_H^4$ at the same order in $1/x_0$ but it's coefficient vanishes; thus there is no contribution coming from the background temperature. Non-zero effects of the background temperature is observed at the next order in $1/x_0$.}
\begin{eqnarray} \label{phiebperpa}
\Delta \phi_{\infty} = \frac{\sqrt{6} \pi}{4} - \sqrt{\frac{3}{2}} \frac{e^2 - h^2}{2 x_0} + \ldots \ .
\end{eqnarray}
The formula in (\ref{phiebperpa}) implies if $e>h$, then asymptotically $\Delta\phi_{\infty} < \sqrt{6} \pi/4$; on the other hand if $e<h$, then asymptotically $\Delta\phi_{\infty} < \sqrt{6} \pi/4$. The first case is similar to the behaviour observed in fig.~\ref{bc@c} and the second case is similar to the behaviour observed in fig.~\ref{bending1}. Thus we expect no phase transition for $e>h$ and any non-trivial phase transition will take place only in the limit $e<h$. These features are pictorially demonstrated in fig.~\ref{pebperp}. We have numerically verified that the qualitative features demonstrated in fig.~\ref{pebperp} are completely generic for both $e/h>1$ and $e/h<1$. 

Alternatively, from the definition of the on-shell action in (\ref{suebperp}) and the asymptotic angle separation obtained from (\ref{phiebperp}) we can obtain
\begin{eqnarray}
&& \frac{S_U}{\cN_T} - \frac{1}{2} c \Delta\phi_\infty  =   \int_{r_0}^{\infty} \frac{1}{r \sqrt{f}} \left( r^8 f Q^2 - 6 c^2 \right)^{1/2} \ ,  \quad Q = \left(1 + \frac{h^2}{r^4} - \frac{e^2}{r^4 f} \right)^{1/2} \ . \\
&& \implies \frac{\partial}{\partial c} \left( \frac{S_U}{\cN_T} - \frac{1}{2} c \Delta\phi_\infty \right)  =  - \frac{1}{2} \Delta \phi_\infty \nonumber\\
&& \implies \frac{1}{\cN_T} \frac{\partial S_U}{\partial c} = \frac{1}{2} c \frac{\partial \Delta\phi_\infty}{\partial c} \ .
\end{eqnarray}
Using the asymptotic expansion in (\ref{phiebperpa}) we get
\begin{eqnarray} \label{suebperpa}
\frac{1}{\cN_T} S_U = {\rm const.} + \sqrt{\frac{3}{2}} \frac{e^2 - h^2}{4} \log c + \ldots 
\end{eqnarray}
From (\ref{suebperp}) it is clear that in the limit $e>h$ the U-shaped embeddings become more and more energetic as $c$ increases; on the other hand, in the limit $e<h$ increasing $c$ decreases the energy of this class of embeddings. Thus we can conclude that for $e>h$, there will be no phase transition since the parallel shaped are always favoured and the interesting physics happens only in the regime where $e<h$.

\begin{figure}[htp]
\begin{center}
{\includegraphics[angle=0,
width=0.65\textwidth]{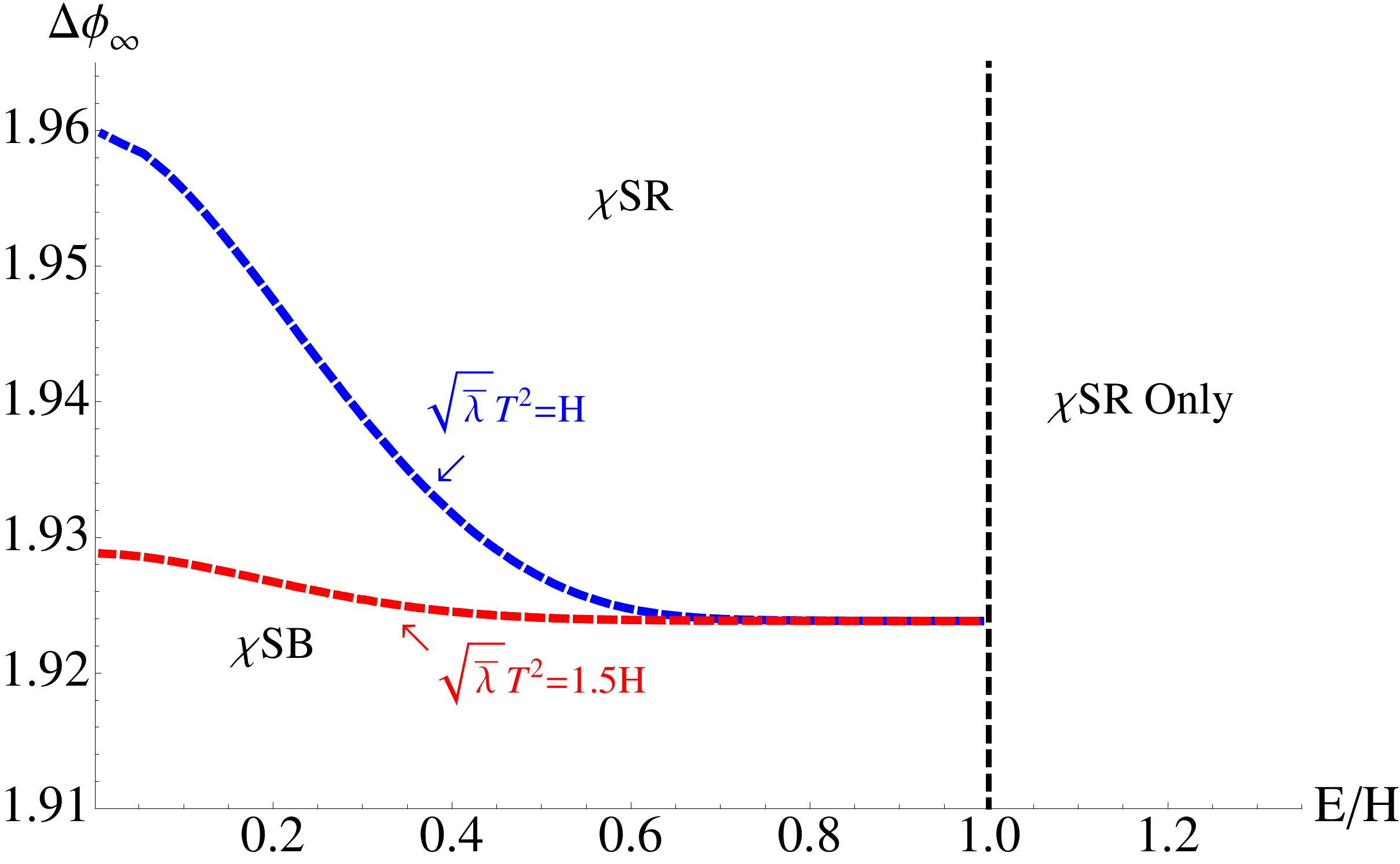} }
\caption{\small The phase diagram in the $(\Delta\phi_{\infty} - E/H)$-plane for various fixed values of the ratio $H/(\sqrt{\bar\lambda} T^2)$. The non-trivial phase structure appears only in the limit $E/H<1$ and in the regime $E>H$ only the chiral symmetry restored phase is available.}
\label{eperpb_1}
\end{center}
\end{figure}
\begin{figure}[htp]
\begin{center}
{\includegraphics[angle=0,
width=0.65\textwidth]{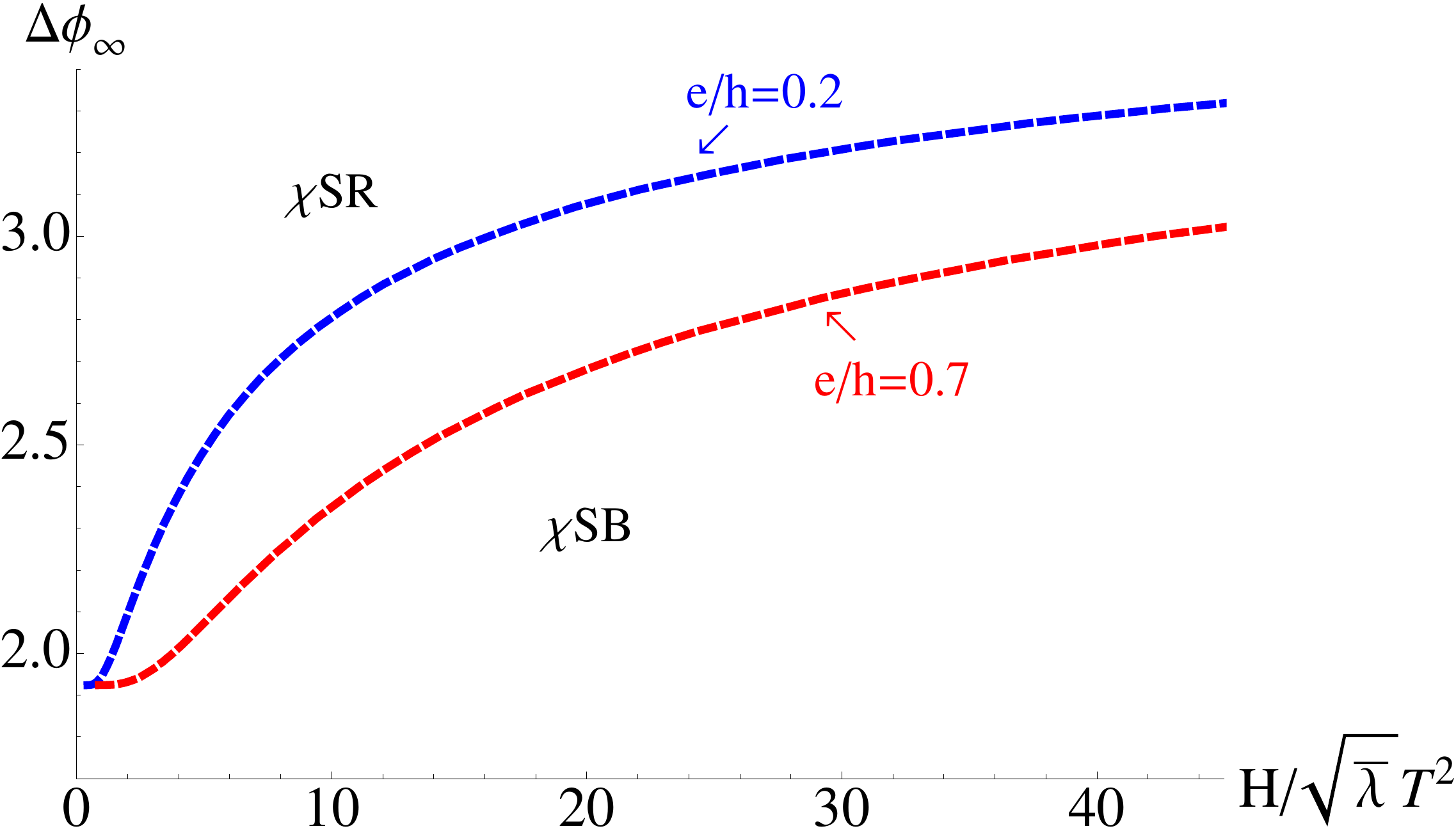} }
\caption{\small The phase diagram in the $\Delta\phi_{\infty} - H/(\sqrt{\bar\lambda} T^2)$-plane for various fixed values of the ratio $E/H$.
Again we restrict ourselves to the regime $E/H<1$.}
\label{eperpb_2}
\end{center}
\end{figure}
We have demonstrated a few representative phase diagrams in $\{ \Delta\phi_{\infty} - E/H \}$-plane $ \{ \Delta\phi_{\infty} - H/(\sqrt{\bar\lambda} T^2) \}$-plane and in fig.~\ref{eperpb_1} and fig.~\ref{eperpb_2} respectively. To avoid clumsiness, we have just presented the corresponding phase boundaries and have not appropriately labeled every region as in fig.~\ref{TBphase}. In fig.~\ref{eperpb_1}, we observe that for a fixed value of $H/(\sqrt{\bar\lambda} T^2)$, increasing electric field decreases the asymptotic angle separation. On the other hand, for a fixed value of $E/H$, increasing $(\sqrt{\bar\lambda} T^2)/H$ decreases $\Delta\phi_\infty$. This behavior is expected since both the electric field and temperature favour the restoration of chiral symmetry, whereas a magnetic field promotes symmetry breaking. Beyond the value $E/H=1$, we do not have any chiral symmetry broken phase at all.

From fig.~\ref{eperpb_2} we observe qualitatively similar physics. For a fixed value of $e/h$, increasing magnetic field increases the asymptotic angle separation, which is in keeping with the effect of magnetic catalysis in chiral symmetry breaking. We also observe that for a given value of $H/(\sqrt{\bar\lambda} T^2)$, increasing $e/h$ decreases $\Delta\phi_\infty$.

\subsection{The case of parallel fields}

For the parallel electric and magnetic fields, the vector potential takes form
\begin{eqnarray}
A_x = - E t + A(r) \ , \quad A_z = H y \ . 
\end{eqnarray}
This time, there is a non-zero Wess--Zumino term, which pushes the  D7 brane away from the equatorial
$\theta=\pi/2$ plane of the $S^2$ sphere.
Consequently, we need to parametrize the brane's geometry by two $r$-dependent angles $\theta( r )$ and $\phi( r )$, or equivalently by a unit 3-vector $\vec n( r )$.
The DBI action is given by
\begin{equation} \label{dbiparallel}
S_{\rm DBI}\ =\ - \cN \int dt dr r^3 \left( 1 + \frac{h^2}{r^4} \right)^{1/2}
\left[ \left(1 - \frac{e^2}{r^4 f}\right) \left( 1 + \frac{r^2}{6} f(r) \vec{n}^{\prime 2} \right)
     + f a'^2 \right]^{1/2} \,,
\end{equation}
where
\begin{equation}
h\  =\  \left( 2 \pi \alpha' H\right) R^2, \quad e\ =\ \left( 2 \pi \alpha' E \right) R^2,\quad
\cN\ =\ \tau_{D7} V_{\mathbb{R}^3} \frac{8 \pi^2}{9} , \quad \frac{h}{r_H^2} = \frac{H}{\sqrt{\bar\lambda} T^2} 
\end{equation}
and
\begin{equation}
\vec n^{\prime 2}\ =\ \theta^{\prime2}\ +\ \sin^2\theta\,\phi^{\prime2}
\end{equation}
(where $'$ denotes $d/dr$),
while the non-zero part of the Wess-Zumino term is
\begin{eqnarray}
S_{\rm WZ} = \frac{\mu_7}{2} \int P\left[\tilde{C}_4\right] \wedge F \wedge F \ , 
\end{eqnarray}
for
\begin{eqnarray}
P\left[\tilde{C}_4\right] = - \frac{R^4}{27 g_s} \cos\theta(r) \phi'(r) f_1 \wedge f_2 \wedge f_3 \wedge dr \ .
\end{eqnarray}
Integrating the WZ term over the 3--space and the $S^3$ gives us
\begin{eqnarray}
S_{\rm WZ} & = & - \frac{eh\mu_7}{27 g_s} \int \cos\theta\phi' dt\wedge dx\wedge dy\wedge dz \wedge dr \wedge f_1 \wedge f_2 \wedge f_3 \nonumber\\
& = & - \frac{2}{3} e h \cN \int dt dr \vec{\cal V}(\vec n)\cdot\vec n'
\label{WZterm}
\end{eqnarray}
where $\vec{\cal V}(\vec n)$ is a vector field on $S^2$ similar to the $\vec A$ field of a magnetic
monopole,
\begin{equation}
\vec{\cal V}(\theta,\phi)\ =\ \cos\theta\,\nabla\phi\
=\ \cot\theta\,\hat\phi\quad
\mbox{($\hat\phi$ is a unit vector in the $\phi$ direction).}
\end{equation}
Altogether, the net Euclidean action for the probe brane takes form
\begin{eqnarray}
S &=& S_{\rm DBI}\ +\ S_{\rm WZ}\ =\ \cN_T \int \! dr\,\cL_{\rm net}\,,\\
\cL_{\rm net} &=& \left( 1 + \frac{h^2}{r^4} \right)^{1/2}
	\left[ \left(1 - \frac{e^2}{r^4 f}\right) \left( 1 + \frac{r^2}{6} f(r) \vec{n}^{\prime 2} \right)\,
     	+\, f a'^2 \right]^{1/2}\
+\ \frac{2}{3} e h \vec{\cal V}\cdot\vec n'.
\end{eqnarray}

For the U-shaped solutions --- where $a\to 0$ for $r\to\infty$ along both sides of the U --- the action is clearly minimized for $a'\equiv 0$.
Consequently, the remaining Lagrangian has form
\begin{eqnarray} \label{lagpara}
\cL = \cA(r) \sqrt{ 1 + \cB(r) \left(\vec{n}\right)^2} + k \vec{\cV}(\vec{n}) \cdot \vec{n}' \ ,
\end{eqnarray}
--- where
\begin{equation} \label{Uparams}
\cA(r)\ =\ r^3 \left( 1 + \frac{h^2}{r^4} \right)^{1/2} \left( 1 - \frac{e^2}{r^4 f(r)} \right)^{1/2} \ , \quad \cB(r)\ =\ \tfrac16\, r^2 f(r) \ , \quad
k\ =\ \tfrac23\, e h \ ,
\end{equation}
--- which resembles Lagrangian of a charged particle moving in magnetic field of a monopole combined with a central electric potential.
As explained in Appendix A, such particle has a modified conserved angular momentum
\begin{equation}
\label{Lmonopole}
{\bf L}\ =\ {\bf r}\times m{\bf v}\ +\ Mq\frac{{\bf r}}{r}
\end{equation}
where $M$ is the magnetic charge of the monopole and $q$ is the electric charge of the particle.
When the ordinary angular momentum ${\bf r}\times m{\bf v}$ is conserved, the particles moves
in a plane $\perp\bf L$.
But for the conserved angular momentum of the form (\ref{Lmonopole}) the particle moves along a cone
making fixed angle  with the $\bf L$ vector.
In spherical coordinates (where $\bf L$ points to the North pole) the radius $r$ and the longitude $\phi$
change with time while the latitude remains constant, $\theta=\rm const\neq\pi/2$.

Likewise, we show in Appendix B that for the U-shaped D7 brane $\theta( r )=\theta_c=\rm const\neq\pi/2$
(in some coordinate system)
while the longitudinal profile $\phi( r )$ depends on the functions $\cA(r)$ and $\cB(r)$.
Specifically,
\begin{equation}
\cos\theta\ \equiv\ {k\over L}\quad{\rm while}\quad
\frac{\cA \cB\,\phi'}{\sqrt{1+\cB(\sin\theta\,\phi')^2}}\ \equiv\ L.
\end{equation}
Assuming the U-shaped brane is smooth at its lowest point $r_0$ (where the D7 brane connects to the
$\overline{\rm D7}$ antibrane), we have $\phi'(r_0)=\infty$, hence
\begin{equation}
\cA(r_0)\sqrt{\cB(r_0)}\ =\ L\sin\theta_c\quad{\rm while}\quad
L\cos\theta_c\ =\ k
\end{equation}
and therefore
\begin{eqnarray}
\label{thetacmain}
\theta_c &=& \arctan\frac{\cA(r_0)\sqrt{\cB(r_0)}}{k} \ , \\
\label{phiebpara}
\frac{d\phi}{dr} &=&
\sqrt{\frac{\cA^2(r_0)\cB(r_0)\,+\,k^2}{[\cA^2(r)\cB(r)\,-\,\cA^2(r_0)\cB(r_0)]\,\cB(r)}} \ .
\end{eqnarray}
Plugging this solution into the Lagrangian (\ref{lagpara}), we obtain the net on-shell
Euclidean action for the U-shaped solution as
\begin{eqnarray}
\label{Uaction}
S_U^E &=& \cN_T\int\limits_{r_0}^\infty\frac{dr}{\sqrt{\cB(r)}}\,
\frac{\cA^2(r)\cB(r)\,+\,k^2}{\sqrt{\cA^2(r)\cB(r)\,-\,\cA^2(r_0)\cB(r_0)}}\\
&=&\cN_T\int\limits_{r_0}^\infty\frac{dr}{r\sqrt{f(r)}}\,
\frac{(r^4+h^2)(fr^4-e^2)\,+\,\frac83 e^2h^2}{\sqrt{(r^4+h^2)(fr^4-e^2)\,-\,(r_0^4+h^2)(f_0r_0^4-e^2)}}\,.
\nonumber
\end{eqnarray}

On the other hand, for the parallel-shaped profile, the equation of motion for the gauge field can be solved to give
\begin{eqnarray}
a' = \frac{j}{r^2 f} \sqrt{\frac{r^4 f - e^2}{\left(r^6 + r^2 h^2 \right)f - j^2}} \ ,
\end{eqnarray}
from which we determine
\begin{eqnarray}
r_*^4 = r_H^4 + e^2 \ , \quad j^2 = \left( r_*^6 + r_*^2 h^2 \right) f(r_*) \ .
\end{eqnarray}
The on-shell Euclidean action for this class of solutions is given by
\begin{eqnarray}
S_{||}^E = \cN_T \int_{r_*}^{\infty} dr \left[ \left( r^4 + h^2 \right) \sqrt{\frac{r^4 f - e^2 }{\left(r^6 + r^2 h^2 \right)f - j^2}} - j a' \right] \ .
\end{eqnarray}
The corresponding phase diagram is obtained by looking at the zeroes of 
\begin{eqnarray}
\Delta S = S_{U}^E - S_{||}^E \ .
\end{eqnarray}

Before proceeding further, let us again investigate the asymptotic angle separation --- or rather the asymptotic longitude separation $\Delta\phi_\infty$ --- in some details. Our goal here is to estimate when a phase transition is possible depending on the relative strength of the electric and the magnetic field. From (\ref{phiebpara}) we obtain:
\begin{eqnarray}
 \Delta\phi_\infty = \frac{3L}{x_0} \int_1^{\infty} \frac{dy}{\sqrt{y \left(y - \frac{r_H^4}{x_0}\right)}} \frac{1}{\left[ \left( y + \frac{h^2}{x_0}\right) \left( y - \frac{r_H^4 + e^2}{x_0}\right) - \left( 1 + \frac{h^2}{x_0}\right) \left( 1 - \frac{r_H^4 + e^2}{x_0}\right) \right]^{1/2}} \ , \nonumber\\ 
\end{eqnarray}
where we have again defined
\begin{eqnarray}
x = r^4 \ , \quad x_0 = r_0^4  \ .
\end{eqnarray}
It can again be shown that in the large $L$ (hence the large $x_0$) limit, the asymptotic longitude separation is given by
\begin{eqnarray}
\Delta\phi_\infty & = & \frac{\sqrt{6} \pi}{4} - \sqrt{\frac{3}{2}} \frac{e^2 - h^2}{2 x_0} + \ldots \ ,
\end{eqnarray}
Interestingly, this is the exact expression we obtained for the perpendicular case in (\ref{phiebperpa}) as well. As before it can also be checked that in this case we get
\begin{eqnarray} \label{suebpara}
\frac{1}{\cN_T}\frac{\partial S_U}{\partial c} = \frac{1}{2} c \frac{\partial \Delta\phi_\infty}{\partial c} + \ldots \ ,
\end{eqnarray}
where we now have
\begin{eqnarray}
c = L \sin^2 \theta_c \ .
\end{eqnarray}
Unlike (\ref{suebperpa}), the above relation in (\ref{suebpara}) holds only in the limit $c \to \infty$. Taking everything together our general conclusion is similar as before: for $e>h$ we will not have any phase transition and chiral symmetry restored phase is the only available phase, but for $e<h$ we will have non-trivial physics and corresponding phase diagrams.

Before presenting the phase diagram a few words about the asymptotic angle separation are in order. Since in this case we have to fix some constant value of $\theta=\theta_c$, which is non-equatorial, the physical angle separation is the three-dimensional one instead of just $\Delta\phi_{\infty}$. If we denote this $3$-d angle separation by $\Delta\Omega$, then in terms of $\theta_c$ and $\Delta\phi_\infty$ this is given by\footnote{This formula is obtained by considering the dot product of two vectors represented by: $x_i = \sin\theta_c \cos\phi_i$, $y_i = \sin\theta_c \sin\phi_i$, $z_i = \cos\theta_c$, with $i=1,2$. Here $\{x, y, z\}$ represent the Cartesian coordinates. Now, taking the dot product we get
\begin{eqnarray}
\cos\Delta\Omega = x_1 x_2 + y_1 y_2 + z_1 z_2 = \cos^2\theta_c + \sin^2 \theta_c \cos \Delta \phi_\infty \ ,
\end{eqnarray}
where $ \Delta \phi_\infty = \phi_1 - \phi_2$.} 
\begin{eqnarray}
\cos\Delta\Omega = \cos^2 \theta_c + \sin^2 \theta_c \cos \Delta\phi_\infty \ .
\end{eqnarray}
The relevant coupling in the dual field theory corresponds to $\Delta\Omega$. We will use $\Delta\Omega$ in the corresponding phase diagrams. 
\begin{figure}[htp]
\begin{center}
{\includegraphics[angle=0,
width=0.65\textwidth]{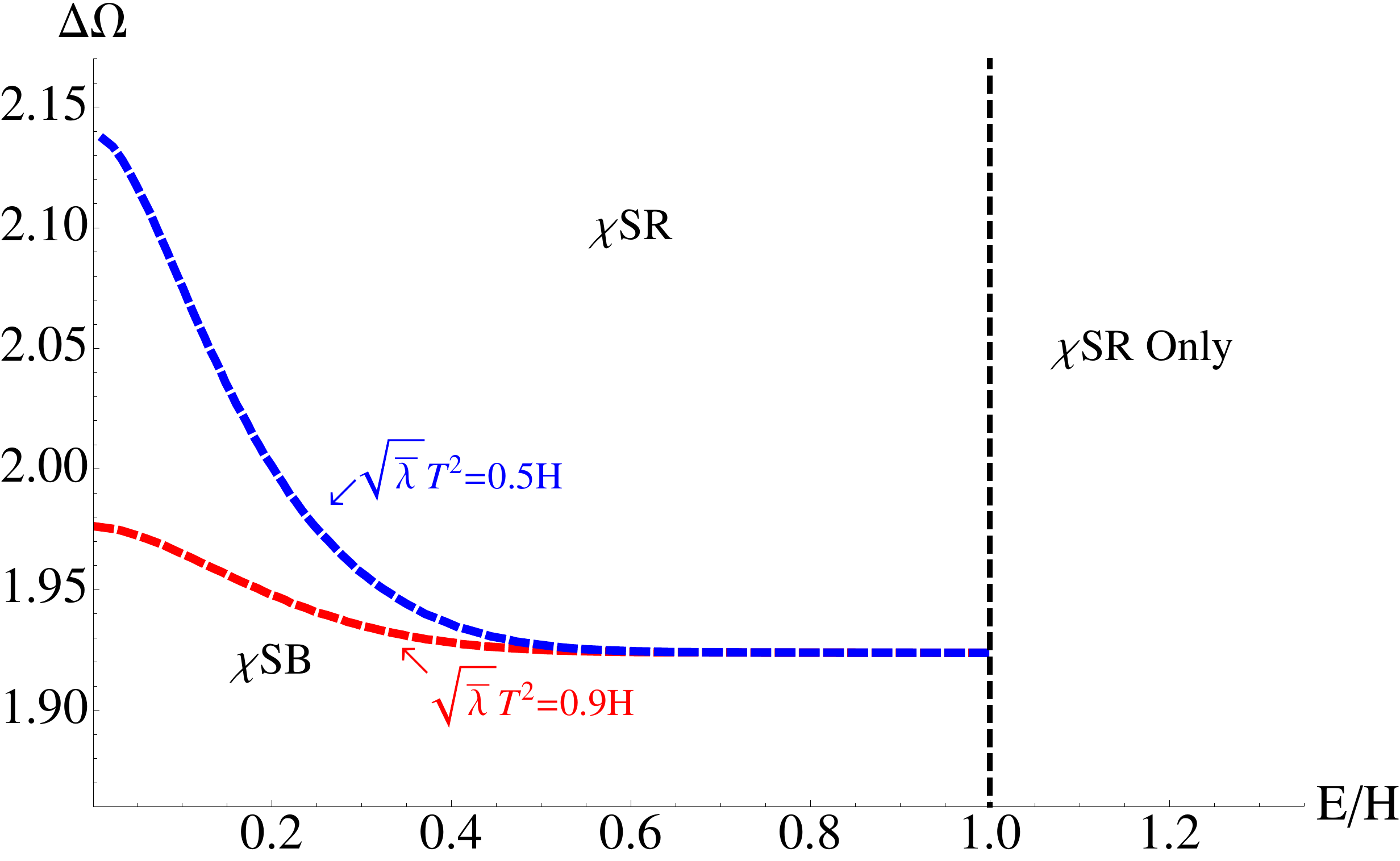} }
\caption{\small The phase diagram in the $(\Delta\Omega - E/H)$-plane for various fixed values of the ratio $(\sqrt{\bar\lambda} T^2)/H$. The non-trivial phase structure appears in the regime $E/H<1$ and in the regime $E/H>1$ only the chiral symmetry restored phase is available.}
\label{eparab_1}
\end{center}
\end{figure}
\begin{figure}[htp]
\begin{center}
{\includegraphics[angle=0,
width=0.65\textwidth]{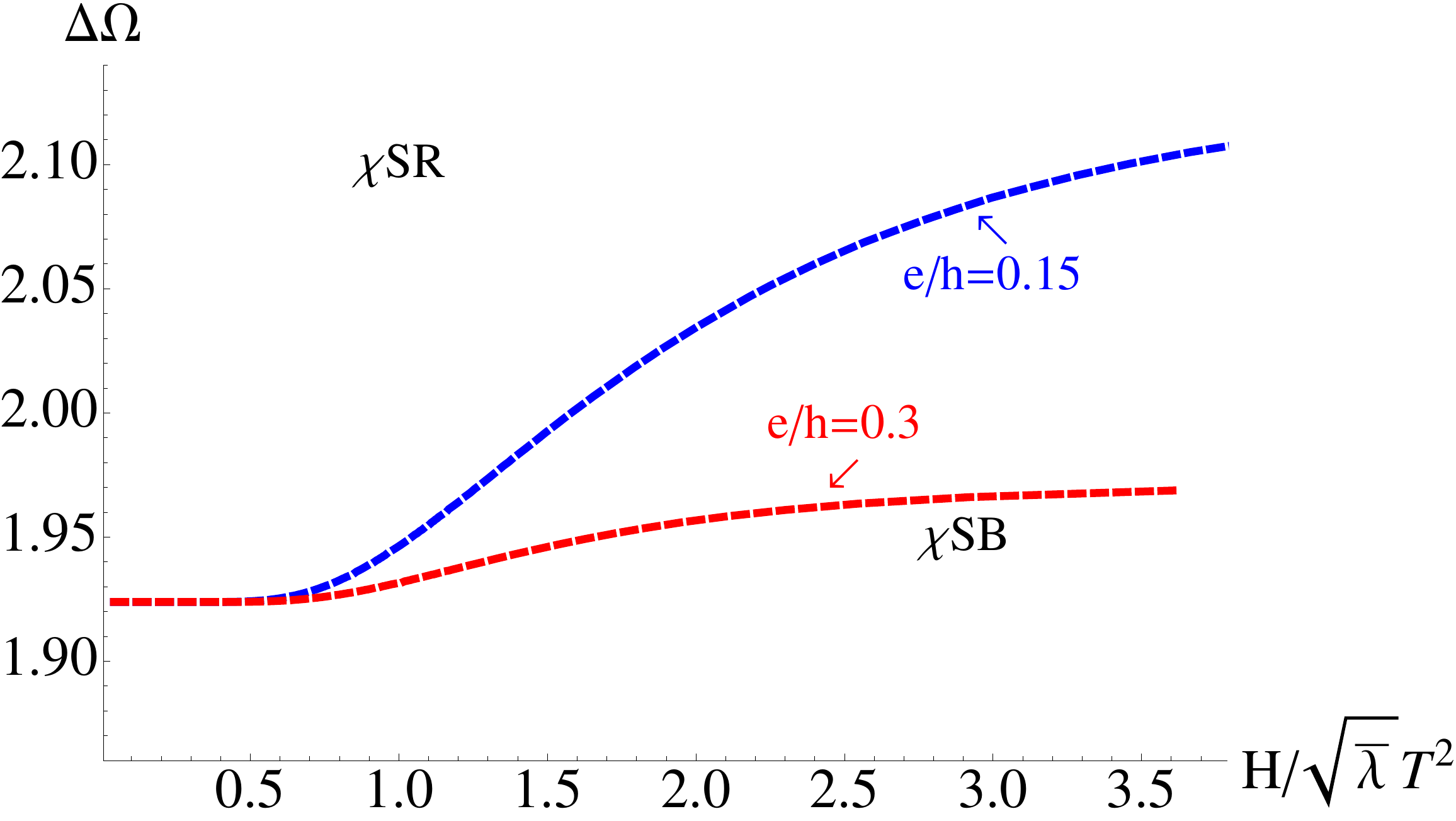} }
\caption{\small The phase diagram in the $\Delta\Omega - H/(\sqrt{\bar\lambda} T^2)$-plane for various fixed values of the ratio $E/H$.
We have restricted ourselves to the interesting regime $E<H$.}
\label{eparab_2}
\end{center}
\end{figure}

A few of the representative phase boundaries has been presented in fig.~\ref{eparab_1} in the $(\Delta\Omega - E/H)$-plane and in fig.~\ref{eparab_2} in the $\{ \Delta\Omega - H/(\sqrt{\bar\lambda} T^2) \}$-plane. As before, in these phase diagrams we have presented just the phase boundaries and have for simplicity. The qualitative features of these phase boundaries are similar to what we have observed for the perpendicular case. For a fixed value of $H/(\sqrt{\bar\lambda} T^2)$, increasing $e/h$ decreases $\Delta\Omega$, where for a fixed value of $e/h$ increasing $H/(\sqrt{\bar\lambda} T^2)$ increases $\Delta\Omega$. As we have already explained before, this is the expected general behavior. 

\newpage

\section{Conclusions}

We have studied in detail the Kuperstein-Sonnenschein model in the presence of temperature and a constant electromagnetic field. We have explicitly demonstrated the effect of {\it magnetic catalysis} in chiral symmetry breaking in this model adding to the claim of this universal phenomenon in various systems. The basic mechanism behind this is the fact that at strong magnetic field the fundamental matter populates the lowest Landau level and the dynamics of flavours effectively becomes $(1+1)$-dimensional.  We have also obtained a detailed phase diagram in the temperature vs magnetic field plane. It is interesting and amusing to compare our phase diagram to the ones obtained earlier from various other holographic models. Such phase diagrams may be relevant for magnetars (neutron stars with a large magnetic field).

Based on our conjectured proposal of free energy in the presence of an electric field, we have also studied in detail the phase diagram when both electric and magnetic fields are present. Two representative situations that we have studied are parallel and perpendicular electric and magnetic fields. Not surprisingly, the qualitative features of these phase diagrams are similar. It is interesting to note that our analysis of a strongly coupled system conforms to weakly coupled field theory intuitions and hence hints towards some robust universal features of such non-Abelian gauge theories. Furthermore, the geometric manifestations of such gauge theory phenomena provide new and interesting outlook.\footnote{For example, within a similar framework a holographic calculation of Schwinger effect has been recently carried out in \cite{Semenoff:2011ng}.}

It is worthwhile to remark here that in what we have considered, the magnetic catalysis (or the effects of the electric field) affects some scalar function which represents the embedding function of the probe brane. Such a bulk scalar field is dual to a single trace fermion bilinear in the dual boundary theory. Recently in \cite{Bolognesi:2011un} the magnetic catalysis of a bulk fermion field has been analyzed, which corresponds to a double trace operator in the boundary theory. Although the construction in \cite{Bolognesi:2011un} is not embedded within string theory, it is interesting to speculate such a possibility and its connection with our work.

There are many avenues for future work. It will be interesting to study the meson spectrum by studying the fluctuations of the probe brane around its classical profile in the presence of these external parameters to supplement our analysis of the various phases. The magnetic field produces interesting effects in the meson spectrum, {\it e.g.} Zeeman splitting, level mixing and enhancing the stability of mesons\cite{Filev:2007gb, Johnson:2009ev} and thus it will be interesting to identify and study such features.

We have not considered the effect of a chemical potential or a non-zero baryon number in this model. A chemical potential can be realized by exciting the time component of the gauge field living on the probe brane worldvolume. The presence of a magnetic field and the baryon number is known to produce novel phases and interesting effects both in the so called ${\rm D3}-{\rm D7}$-model\cite{Evans:2010iy, Jensen:2010vd} and in the Sakai-Sugimoto model\cite{Bergman:2008qv, Thompson:2008qw, Rebhan:2008ur, Rebhan:2009vc}. Thus it will be very interesting to further study the Kuperstein-Sonnenschein model in the presence of a non-zero baryon number.\footnote{Baryon interactions based on non-supersymmetric ${\rm D7}$/$\overline{{\rm D7}}$-brane in the Klebanov-Strassler background has been studied in \cite{Dymarsky:2010ci}.} Having both a magnetic field and a chemical potential gives rise to a non-zero Chern-Simons contribution through an $F \wedge F \wedge C_4$ term. This term has radial dependence through the profile function $\theta(r)$ as well as the gauge field excited on the worldvolume of the probe. This makes the problem technically more challenging as we will no longer be able to consider some constant $\theta$-embedding and will have two coupled differential equations to solve for the profile. Thus we leave this for future work.

Note that one can place the probe ${\rm D7}$-branes in the Klebanov-Witten background in two ways: the supersymmetric embeddings discussed in \cite{Kuperstein:2004hy} and the non-supersymmetric ones which we have studied here. The former embedding does not have any spontaneous chiral symmetry breaking at zero temperature. Moreover, the global ``flavour" symmetry for the supersymmetric embeddings is simply a $U(1)$. These embeddings are in close analogue of the Karch-Katz type embedding of ${\rm D7}$-branes in the ${\rm AdS}_5\times S^5$-background, which we have called the ${\rm D3}-{\rm D7}$-model earlier. It is quite interesting that the Klebanov-Witten background allows for both Karch-Katz type and a Sakai-Sugimoto type construction, although with different global ``flavour" symmetry group and different physics as far as chiral symmetry is considered. Nonetheless, it would be interesting to investigate whether within the Klebanov-Witten background one could perhaps capture the rich phenomenology of both the ${\rm D3}-{\rm D7}$-model and the Sakai-Sugimoto model by considering these two different kinds of ${\rm D7}$-embeddings.

It is noteworthy to remark that the Klebanov-Witten background has a conifold singularity at the origin where both the $S^2$ and the $S^3$ shrinks to zero size. This  singularity is resolved in the Klebanov-Strassler background\cite{Klebanov:2000hb} by considering a deformed conifold and threading an NS-NS three-form flux through the $S^3$. This background corresponds to the confining phase of the dual field theory. Adding non-supersymmetric ${\rm D7}$/$\overline{{\rm D7}}$-branes in this background has been studied in \cite{Dymarsky:2009cm}. To study non-trivial phase diagrams in the presence of external parameters, one needs to know the finite temperature version of the Klebanov-Strassler background which is currently not known in the literature in a closed form. For large temperature, an approximate solution is obtained in \cite{Gubser:2001ri}. It will be interesting to study at least a part of the phase diagram in this large temperature background.

Our analysis is valid in the so called probe limit where the gravitational backreaction of the flavour branes are ignored. One crucial but technically challenging direction is to consider the backreaction of the probe branes and determine the resulting gravitational background. Some earlier works along these lines have been nicely summarized in {\it e.g.} \cite{Nunez:2010sf}. Such an exercise with the non-supersymmetric branes in the Kuperstein-Sonnenschein model could also be a fruitful direction for future work. It is also interesting to speculate how the probe sector physics would be affected in such a back-reacted geometry. Some work along similar directions have been pursued recently in the so called D3--D7 model in \cite{Filev:2011mt, Erdmenger:2011bw}.

Our proposal of the thermodynamic free energy in the presence of the electric field and the boundary current, albeit a physically appealing one, is a conjecture. It would be extremely interesting if it could be directly verified in some simplifying model or argued further. One obvious direction is to consider the backreaction of the probe branes in the presence of the electric field and explicitly demonstrate our claim, which however is a technically difficult task. Some work along similar direction has been carried out in {\it e.g.} \cite{Sahoo:2010sp}.

It is interesting to note that introducing an electric field and consequently having a non-zero current naturally realizes a system out of equilibrium. In the probe limit, according to our proposed definition of thermodynamic free energy, we seem to be able to perform sensible thermodynamics even outside the realm of equilibrium physics. For a system in thermal equilibrium, fluctuation-dissipation theorem relates the fluctuations of the system at equilibrium and its response to applied perturbations. Using our system, we can explore what might be the analogue of such a theorem for systems which are steady-state instead of at strict thermal equilibrium.  Towards this end, we need to analyze the gauge field fluctuations around their classical configurations. We leave this for future investigations.

Finally we conclude by saying that although we do not pretend these models resemble QCD in the microscopic details, many macroscopic (qualitative) properties seem to be universal for these class of strongly coupled large $N_c$ gauge theories (such as the {\it magnetic catalysis}). Thus within this approach we hope to continue to learn interesting and useful lessons relevant to the physics of quarks and gluons.

\section{Acknowledgments}
It is a pleasure to thank Oren Bergman, Anatoly Dymarsky, Gilad Lifschytz and Jacob Sonnenschein for numerous useful discussions and correspondences. We thank Anatoly Dymarsky for collaboration in the early stages of this work. AK would like to thank the hospitality of the Simons Workshop in Mathematics and Physics, the Institute of Nuclear Theory at Seattle and KITP, Santa Barbara during various stages of this work. This material is based upon work supported by the National Science Foundation under Grant PHY--0969020 (all authors) and in part by the National Science Foundation under Grant No. PHY11-25915 (AK), by the Simons Foundation (AK), and by the US--Israeli Bi-National Science Foundation.

\renewcommand{\theequation}{A.\arabic{equation}}
\setcounter{equation}{0}  
\section*{Appendix A. Charged particle in a  monopole field}
\addcontentsline{toc}{section}{Appendix A. Charged particle in a magnetic monopole potential}

In section 7.2 we saw that the effect of the Wess-Zumino term on the motion of the D7 brane
on the $S^2$ is similar to the effect of a monopole magnetic field on the motion of a charged
particle.
So as a warm-up exercise, let us consider the motion of a charged non-relativistic particle
in a magnetic monopole field ${\bf B}=(M/r^2){\bf n}$ superimposed on a central electrostatic
potential $V(r)$.

When the Coulomb electric field of the charged particle is superimposed on the monopole magnetic field,
the Poynting vector $\bf E\times B$ has vorticity and hence non-zero angular momentum
\begin{equation}
{\bf L}_{\rm EM}\ =\int\!d^3{\bf x}\,{\bf x}\times{\bf E\times B}\
=\ -qM\,{\bf n}
\end{equation}
where $q$ is the electric charge of the particle, $M$ is the magnetic charge of the monopole,
and ${\bf n}={\bf r}/r$ is the unit vector in the direction from the monopole to the particle.
Consequently, the net angular momentum of the particle and the EM fields is
\begin{equation}
\label{Lnet}
{\bf L}\ =\ {\bf L}_{\rm particle}\ +\ {\bf L}_{\rm EM}\
=\ {\bf r}\times m{\bf v}\ -\ qM\,{\bf n}\,.
\end{equation}
Note that this is the net angular momentum that is conserved by the particle's motion while
the ${\bf L}_{\rm particle}$ and ${\bf L}_{\rm EM}$ vary due to magnetic torques on the particle.
Indeed,  particle's equation of motion
\begin{equation}
m{\bf a}\ =\ {\bf F}_{\rm net}\ =\ -q\nabla V\ +\ q{\bf v}\times{\bf B}\
=\ -qV'{\bf n}\ +\ {qM\over r^2}\,({\bf v\times n})
\end{equation}
leads to
\begin{equation}
{d\over dt}({\bf r}\times m{\bf v})\ =\ {\bf r}\times{\bf F}_{\rm net}\
=\ 0\ +\ {qM\over r^2}\,{\bf r}\times({\bf v\times n})\
=\ {qM\over r}\,\bigl( {\bf v}\,-\,({\bf vn}){\bf n}\bigr)\
=\ qM\,\frac{d{\bf n}}{dt}
\end{equation}
and hence
\begin{equation}
\frac{d{\bf L}}{dt}\ =\ 0.
\end{equation}

When the ordinary angular momentum ${\bf r}\times m{\bf v}$ is conserved, the particle moves
in the central plane $\perp$ to the  $\bf L$ vector.
In the monopole magnetic field the conserved angular momentum is (\ref{Lnet}), so instead of
a central plane the particle moves along a cone at fixed angle $\theta={\rm const}\neq\pi/2$
to the $\bf L$ vector.
Indeed,
\begin{equation}
{\bf n\cdot L}\ =\ {\bf n}\cdot({\bf r}\times m{\bf v})\ -\ {\bf n}\cdot(qM\,{\bf n})\
=\ 0\ -\ qM\ =\ \rm const
\end{equation}
and therefore
\begin{equation}
\cos\theta\ =\ \frac{{\bf n\cdot L}}{|{\bf L}|}\ =\ -\frac{qM}{L}\ =\ \rm const.
\end{equation}
In the spherical coordinate system with North pole in the $\bf L$ direction,
this constant $\theta$ is the particle's latitude angle, while the particle's motion in the longitudinal
direction $\phi$ is governed by $L$,
\begin{equation}
mr^2{d\phi\over dt}\ =\ L\ =\ \rm const.
\end{equation}
Finally, the radial motion is governed by the energy conservation,
\begin{equation}
E\ =\ {m\over2}\left({dr\over dt}\right)^2\ +\ V(r)\
+\ {L^2\sin^2\theta\,=\,L^2-(qM)^2\over 2mr^2}\ =\ \rm const.
\end{equation}

\renewcommand{\theequation}{B.\arabic{equation}}
\setcounter{equation}{0}  
\section*{Appendix B. Brane Profile for Parallel $E$ and $B$ Fields}
\addcontentsline{toc}{section}{Appendix B. Brane Profile for Parallel $E$ and $B$-Fields}
In section 7.2 we saw that when a D7 brane carries both electric and magnetic fields
that are parallel to each other (or more generally when ${\bf E\cdot B}\neq0$),
the action governing the brane's geometry included a non-trivial
Wess--Zumino term.
Consequently, the path of the brane on the $S^2$ as a function of the radius $r$
does not follow the equator (or any other great circle) but involves both dimensions of
the $S^2$.
In this Appendix we shall see that the brane lies along a latitude circle $\theta=\rm const\neq\pi/2$
and derive its longitudinal profile $\phi(r)$.

The effective action for the profile of a U-shaped brane is spelled out in eq.~(\ref{lagpara}).
In terms of $r$-dependent unit 3-vector $\vec n(r)$,
\begin{equation}
\label{lagparab}
S\ =\int\!dr\Bigl(\cA(r)\sqrt{1+\cB(r)\vec n^{\prime2}}\,+\,k\vec{\cal V}(\vec n)\cdot\vec n'\Bigr)
\end{equation}
where $\vec n'\buildrel{\rm def}\over=d\vec n/dr$,
$\cA$ and $\cB$ are functions of $r$ --- they are spelled out in eq.~(\ref{Uparams}), but their
form is not important for the present argument, --- and $\vec{\cal V}(\vec n)$ is a vector field
on the $S^2$ similar to the $\bf A$ field of a magnetic monopole,
\begin{equation}
\vec{\cal V}(\theta,\phi)\ =\ \cos\theta\,\nabla\phi\ =\ \cot\theta\,\hat\phi,\quad
\nabla\times\vec{\cal V}(\vec n)\ =\ -\vec n \ .
\end{equation}
Indeed, the effect of the WZ term $k\vec{\cal V}\cdot\vec n'$ on the brane profile is similar
to the effect of a magnetic monopole field on the motion of a charged particle discussed
in Appendix~A --- there is an extra $k\vec n$ term in the conserved ($i.\,e.,\ r$-independent)
angular momentum of the brane.

To see how this works, let's develop the analogy between the brane profile $\vec n(r)$
and the particle's motion ${\bf r}(t)$.
For the brane, there is no radial motion, and the radial coordinate $r$ itself plays the
role of time for the angular motion on the $S^2$.
Thus, the first term in the brane's Lagrangian~(\ref{lagparab}) acts as a non-quadratic
$r$-dependent kinetic energy for the angular motion,
hence the analogue of the particle's mechanical momentum ${\bf p}=m{\bf v}$ is
\begin{equation}
\vec P\ =\ \frac{\partial(\mbox{first term in (\ref{lagparab})})}{\partial\vec n' }\
=\ \frac{\cA\cB\,\vec n'}{\sqrt{1+\cB\,\vec n^{\prime2}}}
\end{equation}
while the  canonical momentum is
\begin{equation}
\vec P_{\rm can}\ =\ \frac{\partial{\cal L}}{\partial\vec n'}\
=\ \vec P\ +\ k{\cal V}(\vec n) \ .
\end{equation}
Consequently, the Euler--Lagrange equation for the brane is
\begin{equation}
\frac{d}{dr}\vec P\
=\ \frac{\partial{\cal L}}{\partial\vec n}\ -\ k\,\frac{d\vec{\cal V}(\vec n)}{dr}\
=\ k\,\frac{\partial{\cal V}_j}{\partial\vec n}\,n_j'\ -\ k\,\frac{\partial\vec{\cal V}}{\partial n_j}\,n_j'\
=\ k\,\vec n'\times(\nabla\times\vec{\cal V})\ =\ -k\,\vec n'\times\vec n \ ,
\end{equation}
where the right hand side is analogous to the Lorentz force in a monopole magnetic field.
Finally, the analogy of the conserved net angular momentum is
\begin{equation}
\label{Lbrane}
\vec L\ =\ \vec n\times\vec P\ +\ k\vec n,\quad
\frac{d\vec L}{dr}\ =\ 0.
\end{equation}
Indeed,
\begin{equation}
\frac{d}{dr}(\vec n\times\vec P)\
=\ \vec n'\times\vec P\ +\ \vec n\times\vec P'\
=\ 0\ - \ \vec n\times(k\vec n'\times\vec n)\
=\ - k\vec n'\quad
\Longrightarrow\quad \vec L'\ =\ 0.
\end{equation}

Conservation of the angular momentum (\ref{Lbrane}) containing the $k\vec n$ term leads to
constant angle $\Theta=\rm const$ between the brane and the $\vec L$ vector.
Specifically,
\begin{equation}
\vec n\cdot\vec L\ =\ k\ =\ {\rm const}\quad\Longrightarrow\quad
\cos\Theta\ =\ {k\over|L|}\ =\ {\rm const.}
\end{equation}
Thus, the brane's path on the $S^2$ lies along a circle, but it's not a great circle
since $\cos\Theta\neq0$.\footnote{%
    The brane does follow a great circle $\perp\vec L$ (which can be identified as the equator in some
    coordinate system) when the Wess--Zumino term vanishes, $k=0\ \Longrightarrow\ \cos\Theta=0$.
    This happens when there is only the magnetic field but ${\bf E}=0$, or when there
    is only the electric field but ${\bf B}=0$, or when the $\bf E$ and $\bf B$ fields
    are $\perp$ to each other.
	But when both the electric and the magnetic fields are present and ${\bf E}\not\perp{\bf B}$,
	there is non-zero WZ term $k\propto{\bf E\cdot B}$ which moves the brane away from a great circle
	on the $S^2$, $\cos\Theta\neq 0$.
	}
Instead, we may identify it as a constant-latitude circle
\begin{equation}
\theta(r)\ \equiv\ \Theta\ =\ \arccos{k\over L}\ \neq\ \frac\pi2
\end{equation}
in a spherical coordinate system where the North pole is in the direction of $\vec L$.

As to the motion in the longitudinal direction $\phi(r)$, in 3-vector notations we have
$\vec n'=\vec\omega\times\vec n$ where $\vec\omega$ is a vector of magnitude $\phi'$
pointing due North (same direction as $\vec L$).
Thus,
\begin{eqnarray}
\vec n^{\prime2} &=& (\vec\omega\times\vec n)^2\ =\ \omega^2\sin^2\theta \ ,\\
\vec P &=& \frac{\cA\cB}{\sqrt{1+\cB\,\omega^2\sin^2\theta}}\,\vec\omega\times\vec n \ ,\\
\vec L &=&
\frac{\cA\cB}{\sqrt{1+\cB\omega^2\sin^2\theta}}\,\vec n\times(\vec\omega\times\vec n)\
	+\ k\vec n,\\
&\Downarrow&\nonumber\\
\vec L\ -\ (L\cos\theta)\vec n &=&
\frac{\cA\cB}{\sqrt{1+\cB\,\omega^2\sin^2\theta}}\Bigl(\vec\omega\,-\,(\omega\cos\theta)\vec n\Bigr) \ ,
\end{eqnarray}
and consequently
\begin{equation}
\frac{\cA\cB\,\omega}{\sqrt{1+\cB\,\omega^2\sin^2\theta}}\ \equiv\ L \ .
\end{equation}
Solving this equation for the $\omega=\phi'$, we obtain
\begin{equation}
\frac{d\phi}{dr}\ =\ \frac{L}{\sqrt{\cB(r)}\times\sqrt{\cA^2(r)\cB(r)\,-\,L^2\sin^2\theta}} \ .
\end{equation}

\renewcommand{\theequation}{C.\arabic{equation}}
\setcounter{equation}{0}  
\section*{Appendix C. The Open String Metric}
\addcontentsline{toc}{section}{Appendix C. The Open String Metric}

In \cite{Seiberg:1999vs}, the authors studied open strings in the presence of a constant electromagnetic field. One of the upshots of this is the definition of the open string metric which is different from just the spacetime metric in the presence of such background fields. Let $G$ be the background spacetime metric, $F$ be the constant electromagnetic field; then the open string metric, denoted by $\cS$, is given by
\begin{eqnarray}\label{osm}
&& \cS^{ab} = \left[ \left( G + F \right)^{-1}_{\rm symm}\right]^{ab} = \left[ \left( G + F \right)^{-1} G \left( G - F \right)^{-1} \right]^{ab}\ , \\
&& \cS_{ab} = G_{ab} - \left( F G^{-1} F \right)_{ab} \ , \\
&& \cA^{ab} = \left[\left( G + F \right)^{-1}_{\rm anti-symm}\right]^{ab} = -  \left[ \left( G + F \right)^{-1} F \left( G - F \right)^{-1} \right]^{ab} \ ,
\end{eqnarray}
where $\cA$ is the purely anti-symmetric part. As argued in \cite{Seiberg:1999vs}, the open string metric $\cS$ simply describes the effective metric seen by the open strings.

The ten-dimensional background geometry is given in (\ref{kw}). Following \cite{Kuperstein:2008cq}, we place the D7/anti-D7 brane at $\theta = \pi/2$ and their profile is described by the scalar function $\phi(r)$. With this information, the induced metric on the worldvolume of the probe can be calculated to be given by
\begin{eqnarray} \label{metd7}
ds_{D7}^2 & = & \frac{r^2}{R^2} \left( - f(r) dt^2 + d\vec{x}^2 \right) + \frac{R^2}{r^2 f(r)} \left( 1 + \frac{r^2}{3} (\phi')^2 \right) dr^2  \nonumber\\
&+& \frac{R^2}{3} \left[ \frac{1}{2} \left(f_1^2 + f_2^2 \right) + \frac{1}{3} f_3^2 - \phi' dr f_1\right] \nonumber\\
&=& g_{tt} dt^2 + g_{xx} d\vec{x}^2 + g_{rr} dr^2 + g_{11} f_1^2 + g_{22} f_2^2 + 2 g_{r1} dr f_1 \ .
\end{eqnarray}
Our goal here is to compute the open string metric (for an open string ending on the D7-brane) taking $G$ to be the induced metric in (\ref{metd7}) in the presence of D7-brane worldvolume gauge fields. We consider two particular cases: perpendicular and parallel electric and magnetic field. Not surprisingly, our results will match the corresponding results in \cite{Kim:2011qh}.

\begin{itemize}
\item \noindent {\bf Perpendicular electric and magnetic field:}
\end{itemize}

The world volume gauge field ansatz is given by:
\begin{eqnarray}
A_x = - E t + A(r) \ , \quad A_y = H x \ .
\end{eqnarray}
The open string metric evaluated using the formula in (\ref{osm}) is given by
\begin{eqnarray}
ds^2 & = & \frac{g_{tt} g_{xx} + e^2}{g_{xx}} d\tilde{t}^2 + \left( \frac{e^2}{g_{tt}} + g_{xx} + \frac{h^2}{g_{xx}} + \frac{g_{11} (a')^2}{g_{11} g_{rr} - g_{r1}^2 }\right) dx^2 + \left(g_{xx} \left(g_{tt} g_{xx} + e^2\right) + h^2 g_{tt}\right) d\tilde{y}^2 \nonumber\\
&+& dr^2 \left( g_{rr} + \frac{g_{tt} g_{xx} (a')^2}{g_{xx} \left(g_{tt} g_{xx} + e^2\right) + h^2 g_{tt}}\right) \nonumber\\
&+& g_{xx} dz^2 + g_{11} f_1^2 + g_{22} f_2^2 + g_{33} f_3^2 + g_{r1} dr f_1 \ , \\
d\tilde{t} & = & dt + \frac{e h}{g_{tt} g_{xx} + e^2} dy - \frac{e a'}{g_{tt} g_{xx} + e^2} dr \ , \\
d\tilde{y} & = & \frac{1}{\sqrt{g_{tt} g_{xx} + e^2}} \left( dy + \frac{h a' g_{tt}} {g_{xx} \left(g_{tt} g_{xx} + e^2\right) + h^2 g_{tt}} dr \right) \ .
\end{eqnarray}
Here $g_{\alpha\beta}$ denotes the induced metric on the probe D7/anti-D7 brane. As we have seen before, only the parallel embeddings have a non-trivial $a(r)$. For this class of embeddings, the open string metric has an event horizon (denoted by $r= r_*$) which is different from the event horizon on the induced probe metric. The position of the open string metric event horizon is determined from 
\begin{eqnarray}
&& {\cal S}^{rr} = \left( g_{rr} + \frac{g_{tt} g_{xx} (a')^2}{g_{xx} \left(g_{tt} g_{xx} + e^2\right) + h^2 g_{tt}}\right) ^{-1} = \frac{j^2 + g_{tt} g_{xx}}{g_{tt} g_{rr} g_{xx}^2} = 0 \ , \\
&& \implies j^2 + g_{tt} g_{xx} = 0 \quad \implies \quad g_{tt} \left(g_{xx}^2 + h^2\right) + g_{xx} e^2 = 0 \ .
\end{eqnarray}
where we have used the equation of motion of the gauge field to substitute $a'(r)$ in favour of the constant $j$ and also used the relation from which we fix $j$. It is clear from above that for this class of embeddings the corresponding phase in the dual field theory feels an effective temperature set by the pseudo-horizon. Thus in analogy with the purely finite temperature story, the appropriate ``free energy" in this phase is defined as the on-shell action of the probe D7/anti-D7 brane which goes from $r= r_*$ to $r = \infty$. 

On the other hand, for the U-shaped embeddings we have $a' =0$ and thus the open string metric event horizon and the event horizon on the induced probe brane coincide.

\begin{itemize}
\item {\bf Parallel electric and magnetic field:}
\end{itemize}

For parallel electric and magnetic fields, we take the following ansatz for the gauge fields
\begin{eqnarray}
A_x = - E t + A(r) \ , \quad A_z = H y \ .
\end{eqnarray}
In this case, the open string metric is given by
\begin{eqnarray}
ds^2 & = & \frac{g_{tt} g_{xx} + e^2}{g_{xx}} d\tilde{t}^2 + \frac{g_{xx}^2 + h^2}{g_{xx}} \left(dy^2 + dz^2\right) + \left(\frac{e^2}{g_{tt}} + g_{xx} + \frac{g_{11} (a')^2}{g_{11} g_{rr} - g_{r1}^2}\right) dx^2 \nonumber\\
&+& dr^2 \left( g_{rr} + \frac{g_{tt} (a')^2}{g_{tt} g_{xx} + e^2}\right) + g_{11}f_1^2 + g_{22} f_2^2 + g_{33} f_3^2 + g_{r1} f_1 dr \ , \\
d\tilde{t} & = & dt - \frac{e a'}{g_{tt} g_{xx} + e^2} \ .
\end{eqnarray}
For the parallel embeddings, the open string metric event horizon is determined from the following relation
\begin{eqnarray}
&& {\cal S}^{rr} = \left( g_{rr} + \frac{g_{tt} (a')^2}{g_{tt} g_{xx} + e^2}\right)^{-1} = \left( \frac{g_{rr} g_{tt} \left(g_{xx}^2 + h^2 \right)}{g_{tt} \left(g_{xx}^2 + h^2 \right) + j^2} \right)^{-1} = 0 \ , \nonumber\\
&& \implies \quad g_{tt} \left(g_{xx}^2 + h^2 \right) + j^2 = 0 \quad \implies \quad g_{tt} g_{xx} + e^2 = 0 \ ,
\end{eqnarray}
where we have again used the equation of motion for the gauge field and the relation from which we fix the current $j$. Once again we see the emergence of an effective temperature for the conducting phase which is set by the pseudo-horizon. In this case as well we propose a similar definition of free energy as in the previous section.




\begin{thebibliography}{99}


\bibitem{Maldacena:1997re}
  J.~M.~Maldacena,
  ``The Large N limit of superconformal field theories and supergravity,''
  Adv.\ Theor.\ Math.\ Phys.\  {\bf 2}, 231-252 (1998).
  [hep-th/9711200].
  

\bibitem{Gubser:1998bc}
  S.~S.~Gubser, I.~R.~Klebanov, A.~M.~Polyakov,
  ``Gauge theory correlators from noncritical string theory,''
  Phys.\ Lett.\  {\bf B428}, 105-114 (1998).
  [hep-th/9802109].
  
  
\bibitem{Witten:1998qj}
  E.~Witten,
  ``Anti-de Sitter space and holography,''
  Adv.\ Theor.\ Math.\ Phys.\  {\bf 2}, 253-291 (1998).
  [hep-th/9802150].
  
\bibitem{Aharony:1999ti}
  O.~Aharony, S.~S.~Gubser, J.~M.~Maldacena, H.~Ooguri, Y.~Oz,
  ``Large N field theories, string theory and gravity,''
  Phys.\ Rept.\  {\bf 323}, 183-386 (2000).
  [hep-th/9905111].
  
  
\bibitem{Gubser:2009md}
  S.~S.~Gubser, A.~Karch,
  ``From gauge-string duality to strong interactions: A Pedestrian's Guide,''
  Ann.\ Rev.\ Nucl.\ Part.\ Sci.\  {\bf 59}, 145-168 (2009).
  [arXiv:0901.0935 [hep-th]].
  
  
\bibitem{Karch:2002sh}
  A.~Karch, E.~Katz,
  ``Adding flavor to AdS / CFT,''
  JHEP {\bf 0206}, 043 (2002).
  [hep-th/0205236].
  
  
\bibitem{Babington:2003vm} 
  J.~Babington, J.~Erdmenger, N.~J.~Evans, Z.~Guralnik and I.~Kirsch,
  ``Chiral symmetry breaking and pions in nonsupersymmetric gauge / gravity duals,''
  Phys.\ Rev.\ D {\bf 69}, 066007 (2004)
  [hep-th/0306018].
  

\bibitem{Erdmenger:2007cm} 
  J.~Erdmenger, N.~Evans, I.~Kirsch and E.~Threlfall,
  ``Mesons in Gauge/Gravity Duals - A Review,''
  Eur.\ Phys.\ J.\ A {\bf 35}, 81 (2008)
  [arXiv:0711.4467 [hep-th]].
  
  
\bibitem{Sakai:2004cn}
  T.~Sakai, S.~Sugimoto,
  ``Low energy hadron physics in holographic QCD,''
  Prog.\ Theor.\ Phys.\  {\bf 113}, 843-882 (2005).
  [hep-th/0412141].
  
  
\bibitem{Sakai:2005yt}
  T.~Sakai, S.~Sugimoto,
  ``More on a holographic dual of QCD,''
  Prog.\ Theor.\ Phys.\  {\bf 114}, 1083-1118 (2005).
  [hep-th/0507073].
  
  
\bibitem{Kuperstein:2008cq}
  S.~Kuperstein and J.~Sonnenschein,
  ``A New Holographic Model of Chiral Symmetry Breaking,''
  JHEP {\bf 0809}, 012 (2008)
  [arXiv:0807.2897 [hep-th]].
  
  
\bibitem{Klebanov:1998hh}
  I.~R.~Klebanov, E.~Witten,
  ``Superconformal field theory on three-branes at a Calabi-Yau singularity,''
  Nucl.\ Phys.\  {\bf B536}, 199-218 (1998).
  [hep-th/9807080].


\bibitem{Filev:2007gb}
  V.~G.~Filev, C.~V.~Johnson, R.~C.~Rashkov, K.~S.~Viswanathan,
  ``Flavoured large N gauge theory in an external magnetic field,''
  JHEP {\bf 0710}, 019 (2007).
  [hep-th/0701001].
  

\bibitem{Albash:2007bk}
  T.~Albash, V.~G.~Filev, C.~V.~Johnson, A.~Kundu,
  ``Finite temperature large N gauge theory with quarks in an external magnetic field,''
  JHEP {\bf 0807}, 080 (2008).
  [arXiv:0709.1547 [hep-th]].
  
  
\bibitem{Albash:2007bq}
  T.~Albash, V.~G.~Filev, C.~V.~Johnson, A.~Kundu,
  ``Quarks in an external electric field in finite temperature large N gauge theory,''
  JHEP {\bf 0808}, 092 (2008).
  [arXiv:0709.1554 [hep-th]].
  
  
\bibitem{Erdmenger:2007bn}
  J.~Erdmenger, R.~Meyer, J.~P.~Shock,
  ``AdS/CFT with flavour in electric and magnetic Kalb-Ramond fields,''
  JHEP {\bf 0712}, 091 (2007).
  [arXiv:0709.1551 [hep-th]].
  
  
\bibitem{Evans:2010xs} 
  N.~Evans, T.~Kalaydzhyan, K.~-y.~Kim and I.~Kirsch,
  ``Non-equilibrium physics at a holographic chiral phase transition,''
  JHEP {\bf 1101}, 050 (2011)
  [arXiv:1011.2519 [hep-th]].
  
  
\bibitem{Bergman:2008sg}
  O.~Bergman, G.~Lifschytz, M.~Lippert,
  ``Response of Holographic QCD to Electric and Magnetic Fields,''
  JHEP {\bf 0805}, 007 (2008).
  [arXiv:0802.3720 [hep-th]].
  
  
\bibitem{Johnson:2008vna}
  C.~V.~Johnson, A.~Kundu,
  ``External Fields and Chiral Symmetry Breaking in the Sakai-Sugimoto Model,''
  JHEP {\bf 0812}, 053 (2008).
  [arXiv:0803.0038 [hep-th]].  
  
  
\bibitem{Kundu:2010ye}
  A.~Kundu,
  ``External Fields and the Dynamics of Fundamental Flavours in Holographic Duals of Large N Gauge Theories,''
  [arXiv:1012.5450 [hep-th]].
  
  
\bibitem{Evans:2011mu}
  N.~Evans, A.~Gebauer, K.~-Y.~Kim,
  ``E, B, $\mu$, T Phase Structure of the D3/D7 Holographic Dual,''
  JHEP {\bf 1105}, 067 (2011).
  [arXiv:1103.5627 [hep-th]].
  
\bibitem{Evans:2011tk}
  N.~Evans, K.~-Y.~Kim, J.~P.~Shock,
  ``Chiral phase transitions and quantum critical points of the D3/D7(D5) system with mutually perpendicular E and B fields at finite temperature and density,''
  JHEP {\bf 1109}, 021 (2011).
  [arXiv:1107.5053 [hep-th]].
  
  
\bibitem{Gusynin:1995nb}
  V.~P.~Gusynin, V.~A.~Miransky, I.~A.~Shovkovy,
  ``Dimensional reduction and catalysis of dynamical symmetry breaking by a magnetic field,''
  Nucl.\ Phys.\  {\bf B462}, 249-290 (1996).
  [hep-ph/9509320].
  
  
\bibitem{Semenoff:1999xv}
  G.~W.~Semenoff, I.~A.~Shovkovy, L.~C.~R.~Wijewardhana,
  ``Universality and the magnetic catalysis of chiral symmetry breaking,''
  Phys.\ Rev.\  {\bf D60}, 105024 (1999).
  [hep-th/9905116].
  
  
\bibitem{Miransky:2002eb}
  V.~A.~Miransky,
  ``Dynamics of QCD in a strong magnetic field,''
  [hep-ph/0208180].
  
  
\bibitem{Karch:2007pd}
  A.~Karch, A.~O'Bannon,
  ``Metallic AdS/CFT,''
  JHEP {\bf 0709}, 024 (2007).
  [arXiv:0705.3870 [hep-th]].
  
  
\bibitem{Albash:2006bs}
  T.~Albash, V.~G.~Filev, C.~V.~Johnson, A.~Kundu,
  ``Global Currents, Phase Transitions, and Chiral Symmetry Breaking in Large N(c) Gauge Theory,''
  JHEP {\bf 0812}, 033 (2008).
  [arXiv:hep-th/0605175 [hep-th]].
  
  
\bibitem{Das:2010yw}
  S.~R.~Das, T.~Nishioka, T.~Takayanagi,
  ``Probe Branes, Time-dependent Couplings and Thermalization in AdS/CFT,''
  JHEP {\bf 1007}, 071 (2010).
  [arXiv:1005.3348 [hep-th]].
  
  
\bibitem{Romans:1984an}
  L.~J.~Romans,
  ``NEW COMPACTIFICATIONS OF CHIRAL N=2 d = 10 SUPERGRAVITY,''
  Phys.\ Lett.\  {\bf B153}, 392 (1985).  
  

\bibitem{Karch:2010kt}
  A.~Karch, S.~L.~Sondhi,
  ``Non-linear, Finite Frequency Quantum Critical Transport from AdS/CFT,''
  JHEP {\bf 1101}, 149 (2011).
  [arXiv:1008.4134 [cond-mat.str-el]].
  
  
\bibitem{Karch:2008uy}
  A.~Karch, A.~O'Bannon and E.~Thompson,
  ``The Stress-Energy Tensor of Flavor Fields from AdS/CFT,''
  JHEP {\bf 0904}, 021 (2009)
  [arXiv:0812.3629 [hep-th]].
  
\bibitem{Seiberg:1999vs}
  N.~Seiberg, E.~Witten,
  ``String theory and noncommutative geometry,''
  JHEP {\bf 9909}, 032 (1999).
  [hep-th/9908142].
  
  
\bibitem{Kim:2011qh}
  K.~-Y.~Kim, J.~P.~Shock, J.~Tarrio,
  ``The open string membrane paradigm with external electromagnetic fields,''
  JHEP {\bf 1106}, 017 (2011).
  [arXiv:1103.4581 [hep-th]].
  
  
\bibitem{O'Bannon:2007in}
  A.~O'Bannon,
  ``Hall Conductivity of Flavor Fields from AdS/CFT,''
  Phys.\ Rev.\  {\bf D76}, 086007 (2007).
  [arXiv:0708.1994 [hep-th]].
  
  
\bibitem{Semenoff:2011ng}
  G.~W.~Semenoff, K.~Zarembo,
  ``Holographic Schwinger Effect,''
  [arXiv:1109.2920 [hep-th]].
  
  
\bibitem{Bolognesi:2011un}
  S.~Bolognesi, D.~Tong,
  ``Magnetic Catalysis in AdS4,''
[arXiv:1110.5902 [hep-th]].

  

\bibitem{Johnson:2009ev}
  C.~V.~Johnson, A.~Kundu,
  ``Meson Spectra and Magnetic Fields in the Sakai-Sugimoto Model,''
  JHEP {\bf 0907}, 103 (2009).
  [arXiv:0904.4320 [hep-th]].
  
  
\bibitem{Evans:2010iy}
  N.~Evans, A.~Gebauer, K.~-Y.~Kim, M.~Magou,
  ``Holographic Description of the Phase Diagram of a Chiral Symmetry Breaking Gauge Theory,''
  JHEP {\bf 1003}, 132 (2010).
  [arXiv:1002.1885 [hep-th]].
  
  
\bibitem{Jensen:2010vd}
  K.~Jensen, A.~Karch, E.~G.~Thompson,
  ``A Holographic Quantum Critical Point at Finite Magnetic Field and Finite Density,''
  JHEP {\bf 1005}, 015 (2010).
  [arXiv:1002.2447 [hep-th]].
  
   
  
\bibitem{Bergman:2008qv}
  O.~Bergman, G.~Lifschytz, M.~Lippert,
  ``Magnetic properties of dense holographic QCD,''
  Phys.\ Rev.\  {\bf D79}, 105024 (2009).
  [arXiv:0806.0366 [hep-th]].
  
 
\bibitem{Thompson:2008qw}
  E.~G.~Thompson, D.~T.~Son,
  ``Magnetized baryonic matter in holographic QCD,''
  Phys.\ Rev.\  {\bf D78}, 066007 (2008).
  [arXiv:0806.0367 [hep-th]].
  
  
\bibitem{Rebhan:2008ur}
  A.~Rebhan, A.~Schmitt, S.~A.~Stricker,
  ``Meson supercurrents and the Meissner effect in the Sakai-Sugimoto model,''
  JHEP {\bf 0905}, 084 (2009).
  [arXiv:0811.3533 [hep-th]].
  
  
\bibitem{Rebhan:2009vc}
  A.~Rebhan, A.~Schmitt, S.~A.~Stricker,
  ``Anomalies and the chiral magnetic effect in the Sakai-Sugimoto model,''
  JHEP {\bf 1001}, 026 (2010).
  [arXiv:0909.4782 [hep-th]].
  
  
\bibitem{Dymarsky:2010ci}
  A.~Dymarsky, D.~Melnikov, J.~Sonnenschein,
  ``Attractive Holographic Baryons,''
   [arXiv:1012.1616 [hep-th]].
   

\bibitem{Kuperstein:2004hy}
  S.~Kuperstein,
  ``Meson spectroscopy from holomorphic probes on the warped deformed conifold,''
  JHEP {\bf 0503}, 014 (2005).
  [hep-th/0411097].
  
  
\bibitem{Klebanov:2000hb}
  I.~R.~Klebanov, M.~J.~Strassler,
  ``Supergravity and a confining gauge theory: Duality cascades and chi SB resolution of naked singularities,''
  JHEP {\bf 0008}, 052 (2000).
  [hep-th/0007191].
   
  
\bibitem{Dymarsky:2009cm}
  A.~Dymarsky, S.~Kuperstein, J.~Sonnenschein,
  ``Chiral Symmetry Breaking with non-SUSY D7-branes in ISD backgrounds,''
  JHEP {\bf 0908}, 005 (2009).
  [arXiv:0904.0988 [hep-th]].
  
  
\bibitem{Gubser:2001ri}
  S.~S.~Gubser, C.~P.~Herzog, I.~R.~Klebanov, A.~A.~Tseytlin,
  ``Restoration of chiral symmetry: A Supergravity perspective,''
  JHEP {\bf 0105}, 028 (2001).
  [hep-th/0102172].
  
  
\bibitem{Nunez:2010sf}
  C.~Nunez, A.~Paredes, A.~V.~Ramallo,
  ``Unquenched flavor in the gauge/gravity correspondence,''
  Adv.\ High Energy Phys.\  {\bf 2010}, 196714 (2010).
  [arXiv:1002.1088 [hep-th]].
  

\bibitem{Filev:2011mt} 
  V.~G.~Filev and D.~Zoakos,
  ``Towards Unquenched Holographic Magnetic Catalysis,''
  JHEP {\bf 1108}, 022 (2011)
  [arXiv:1106.1330 [hep-th]].
  
  
\bibitem{Erdmenger:2011bw} 
  J.~Erdmenger, V.~G.~Filev and D.~Zoakos,
  ``Magnetic catalysis with massive dynamical flavours,''
  arXiv:1112.4807 [hep-th].
  
  
\bibitem{Sahoo:2010sp}
  B.~Sahoo, H.~-U.~Yee,
  ``Electrified plasma in AdS/CFT correspondence,''
  JHEP {\bf 1011}, 095 (2010).
  [arXiv:1004.3541 [hep-th]].
  
   

\end{thebibliography}
\end{document}